\documentclass[a4paper,11pt]{article}
\usepackage{jheppub} 
\usepackage{lineno}
\usepackage{subcaption}

\title{\boldmath Using the Virasoro conditions to revise the model of a quark-gluon string fragmentation\footnote{The work was performed with the support of the Ministry of Science and Higher Education of the Russian Federation, project ``Fundamental and applied research of cosmic rays'', No. FSWU-2023-0068.}}

\author{R. V. Nikolaenko}
\affiliation{National Research Nuclear University MEPhI (Moscow Engineering Physics Institute),\\
Kashirskoe shosse 31, 115409 Moscow, Russia}

\emailAdd{rvnikolaenko@mephi.ru}

\abstract{Quark-gluon string fragmentation is usually seen as a universal process with its physics remaining unchanged for different colliding systems and center-of-mass energy. However, there is a way to go beyond this assumption by considering the angular momentum of the fragmenting system. An accurate calculation of the spin of the string requires a specific definition of the initial state of the string. In this paper, I present a new method for defining the initial data for the Nambu-Goto relativistic string and derive the basics of its fragmentation process using the so-called Virasoro conditions. It is shown that very non-trivial consequences arise for non-zero mass strings, including the discrete spectrum of string spin, the discreteness of the possible string breaking points, the close-to-Regge behavior of the spin-mass relation for the light string fragments, and the natural stopping mechanism for fragmentation.}

\begin{document}
\maketitle
\flushbottom

\section{Introduction}
\label{sec:intro}
Hadronic interactions at high energies produce a large number of partons (quarks, antiquarks, diquarks, antidiquarks and gluons) in multiple elementary interactions and time- and space-like parton showers. A key role to understanding the nature of the collisions lies in hadronization, the process of transition from formed partons to stable particles (hadrons) recorded by the detectors. The set of partons produced as a result of a hard process or multiple soft processes does not fully determine the spectrum of final particles. So, the hadronization mechanism can significantly affect the predictions of the hadronic interaction model.

The present work was inspired by the problem of the excess of muons in the extensive air showers of very high energies compared to the predictions of hadronic interaction models, the so-called ``muon puzzle'' \cite{PETRUKHIN2014228, Dembinski_2019, Petrukhin2021, Albrecht_2022, Petrukhin2022}. Several studies are dedicated to finding the solution to this discrepancy between experimental data and theoretical expectations; see \cite{Dembinski_2019, Albrecht_2022} for a comprehensive review. Although some improvements are seen after the re-tuning of hadronic interaction models according to the latest accelerator data \cite{epos_7, sibyll23d, qgsjetiii_1, qgsjetiii_2} and with taking into account the collective effects (usually associated with Quark-Gluon Plasma formation or string interactions in dense medium), these changes alone do not seem to resolve the problem completely. Thus, the physics of the regular string hadronization is again of interest.

Up to this date, the string model of hadron production is the most consistent and reliable method to describe the hadronization process. The actual implementations can be divided into two main approaches: the Lund string model \cite{sclModGlJets, GenModStrFrag} that was embodied in the commonly used Monte-Carlo generator Pythia \cite{Pyth_man, Pyth_phys}, as well as the Nambu-Goto relativistic string-based approach originally proposed for the Caltech-II model \cite{CltchII_2, CltchII_1, CltchII_3} and used in modern hadronic interaction models such as NEXUS \cite{PBGRFT, neXus, neXus_2, neXus_3, neXus_4} and EPOS \cite{epos_1, epos_2, epos_3, epos_4, epos_5, epos_6, epos_7, epos_8} (``corona'' part \cite{core_corona_1, core_corona_2}). Although it can be said that, without a doubt, these models turned out to be extremely successful in the description of a broad range of experimental data, the theoretical foundations, nevertheless, exhibit room for further improvement.

A common characteristic of existing string models is their universality in relation to the properties of the colliding system \cite{qgsjetiii_2}. In order to solve this issue, different ways of adding collective effects into the fragmentation process were proposed \cite{ropes_1, ropes_2, ropes_3, collstr, strshov}.  An interesting alternative way of breaking the universality of string fragmentation is to implement an angular momentum conservation mechanism. While the complexity of the proper quantum theory-based approach to angular momentum conservation is of no discussion, at least a classical model interpretation may be used to take into account particular properties of the interaction. The early string hadronization models were developed and tuned according to the data from $e^+e^-$ collisions (see, for example, \cite{fragmStat, LEPtune}), where angular momentum of any final parton system is mostly (but not entirely, however) defined by the spin properties of the particles at the string end-points. On the contrary, in hadronic collisions an impact parameter of the interacting partons may produce a huge rotation momentum between the string end-points. This property alone can influence the particle production during string fragmentation, as was shown in ref. \cite{nikolaenko}.

The Nambu-Goto string apparatus may be used to calculate the spin-related properties of the string. Thus, the first task is to develop a method for proper defining of the initial state of the relativistic string with spin and non-zero mass.

To avoid confusion, the adopted terminology should first be clarified. In the string theory, the term ``massless'' is often used to mark the absence of the mass term in the string action (e.g. the Nambu-Goto action). However, throughout the course of this paper, by the word ``massless'' we will understand the string with zero invariant mass, i.e. $M_{\text{inv}}^2 \equiv P^2 \equiv P_{\mu}P^{\mu} = 0$, where $P_{\mu}$ ($\mu = 0,~\ldots,~3$) is the total momentum 4-vector of the string. Consequently, the ``massive'' string will refer to a string with $M_{\text{inv}} \ne 0$, i.e. with non-zero center-of-mass energy. In addition, only the string with free ends will be considered in this paper. Although taking into account the mass of heavy quarks at the string ends is essential for heavy-flavor fragmentation \cite{Bowler1981, Morris_bowler}, it significantly magnifies the complexity of the problem and, thus, is not considered here.

It might seem excessive to initiate a reconsideration of such a well-known case. However, the reason can be found after a closer investigation of the apparatus of existing string hadronization models. String models of hadron production stay in the domain of classical theory and require an explicit calculation of the string motion. This means that one has to define the string with given properties (like energy, mass, momentum, position, etc.) obtained from the parton level generator. Thus, the adopted ansatz for how the string is initially defined is of fundamental importance for the model.

The Caltech-II model, for example, uses string equations of motion derived from the Nambu-Goto action. To define the string, the piece-wise constant functions are used \cite{CltchII_1}. It turns out, though, that such an ansatz does not satisfy the Virasoro conditions if the string is massive (see Section 3 for detailed explanation). The Lund model is based on the very simple equation of string motion, where two ends of the string fly apart until all their energy is transferred to the confining field between them \cite{GenModStrFrag}. As the string dynamics in Lund model is not derived from the first principles, it is unclear whether there are similar additional restrictions for Lund string as the Virasoro conditions for the Nambu-Goto string. But it is interesting to see if it is possible to mimic the Lund picture using the properly defined Nambu-Goto string. Moreover, the influence of the restrictions must also be studied for the daughter strings. Thus, a development of the string hadronization model with Virasoro conditions taken into account is required.

The first goal of this article is to present a reliable and consistent method for defining a massive quark-gluon string with spin. This method is based on defining the initial conditions for the Cauchy problem for the string motion as a finite-order eigenfunction expansion (FOEE), so that all restrictions arising from the strict theory of the relativistic string, including the Virasoro conditions, were satisfied. The particular focus of this method is to define the relativistic string for hadronization modeling. Therefore, the formulae for the string coordinates, velocity, and invariant area are the main results presented here.

The second goal is to derive the mathematical basics for describing the fragmentation process of this string. The secondary (daughter) strings should also be correctly defined, and this requirement drastically changes the picture of the fragmentation process. Apart from being defined in a mathematically correct way, string fragmentation must result in a realistic picture of hadron production. Because of that, a new type of fragmentation mechanism with energy release is required. The final fragmentation scheme features several essential properties that are considered in detail.

In Section \ref{sec:NGmodel_base}, I review the basics of the string model of hadronization and the Nambu-Goto relativistic string theory. In Section \ref{sec:ICprob}, the problem of defining the initial conditions for the non-zero mass string is demonstrated. In Section \ref{sec:FOEEmethod}, the concept of the FOEE method is described. Section \ref{sec:FOEE_1} is dedicated to the derivation of the algorithm to calculate the arbitrary moving string within the context of a special case of the FOEE-defined string, which I call an FOEE(1)-string. Examples of string motion calculations are also shown there. In Section \ref{sec:fragm}, I derive the necessary formulae for the description of string fragmentation. The statement of the boundary-value problem for the daughter strings is formulated. The results of imposing the Virasoro conditions on the daughter strings are considered. A new fragmentation mechanism is proposed that allows the release of energy with its redistribution between the daughters.  In Section \ref{sec:discuss}, the results obtained are discussed, and some other possibilities for model improvement are considered.

\section{Main principles of the Nambu-Goto string hadronization model}
\label{sec:NGmodel_base}
Let us briefly present the main provisions of the string model. It is reasonable to assume that already in the early stages after the formation of new partons as a result of multiple elementary interactions occurring in hadronic collisions, these partons combine into massive colorless systems (color singlets) \cite{preconf}. This is the essence of the so-called preconfinement: although at high energies (or at small distances) quarks move quasi-freely relative to each other, they are, nevertheless, causally connected to each other by fields generated by their color charges. The QCD field between quarks, gluons and antiquarks is stretched into a tube (flux tube) as they move away from each other \cite{Kaidalov_1, Kaidalov_2}. The reason for this is the presence of vacuum fields in QCD, which exert external pressure on the field between quarks. This configuration of the QCD field between two partons is also confirmed by the lattice QCD calculations \cite{str_lattice}. It is worth to note that the transverse size of the color field tube is usually neglected in models in order to obtain a one-dimensional object -- a string. This makes it possible to use the mathematical apparatus developed in string theory. However, there are many indications of the need to take into account the transverse expansion of the field for describing the interaction of strings in a dense medium. So, numerous studies consider the effects of the so-called rope formation \cite{ropes_1, ropes_2, ropes_3} and overlapping strings \cite{collstr, strshov}. The models of interacting strings were initially used for studies of high-energy nucleus collisions \cite{ropes_app1, ropes_app2, ropes_app3, ropes_app4, ropes_app5, ropes_app6, ropes_app7}, though in the modern days they are also applied to the $pp$ collisions at highest energies. These models are encouraged by the observed strangeness enhancement \cite{strEnch_2, strEnch_1, strEnch_3}.

As the partons continue to fly apart, the transition from one bound state to a system of two becomes energetically favorable. To model this process, an analogue of the Schwinger mechanism in QED \cite{schwinger_1, schwinger_2, schwinger_3, schwinger_4} applied for the QCD field is used \cite{schwinger_5, schwinger_6, schwinger_7, schwinger_8, schwinger_9, schwinger_10, schwinger_11}. A parton-antiparton pair from the vacuum inside the QCD field tube tunnels through the potential, and the string breaks (fragments) into two. In this case, one of the newly formed partons joins the first fragment, and the other one joins the second fragment (the QCD field lines must always originate from and arrive at color charges). Hadronization is, thus, an iterative process of fragmentation of quark-gluon strings with sufficiently light strings being identified with hadrons (a stable system) and not fragmenting. In this model, mesons are obtained from quark-antiquark strings ($q-\overline{q}$ system), baryons are from quark-diquark strings ($q-qq$ system), and antibaryons come from antiquark-antidiquark strings ($\overline{q}-\overline{qq}$).

Artru and Mennessier \cite{AreaDec} proposed a simple law for string fragmentation: it assigns a constant probability $P_0$ of string break per unit area of the world sheet (the world sheet is the surface that a string sweeps behind it as it moves through space-time):
\begin{equation}
    \label{eq:AreaDec}
    \mathrm{d}P_{\text{break}}=P_0 \mathrm{d}A.
\end{equation}
This law, which is usually called the area decay law, is a generalization of the law of radioactive decay, where the probability of decay of a point-like particle is proportional to the interval of its proper time. But here the role of proper time is played by the invariant area $A$.

To calculate the invariant area of the string and the characteristics of the string fragments, it is necessary to derive equations that describe the dynamics of the string. The Nambu-Goto action for a relativistic string has the form \cite{NGact_1, NGact_2, NGact_3, NGact_4}:
\begin{equation}
    \label{eq:NGaction}
    S_{\text{string}}=-\kappa \int_{\sigma_1}^{\sigma_2}d\sigma \int_{\tau_1(\sigma)}^{\tau_2(\sigma)} d\tau \sqrt{{(\dot{x}x^{\prime})}^2-{\dot{x}}^2{x^{\prime}}^{2}},
\end{equation}
where
\begin{equation*}
    \dot{x}_{\mu}(\tau,\sigma) \equiv \frac{\partial x_{\mu}(\tau,\sigma)}{\partial\tau},\quad x^{\prime}_{\mu}(\tau,\sigma) \equiv \frac{\partial x_{\mu}(\tau,\sigma)}{\partial\sigma}.
\end{equation*}
Here, $x_{\mu}(\tau,\sigma)$ ($\mu = 0, ~\ldots,~3$) is a 4-dimensional vector of string coordinates, $\tau$ is an evolutionary parameter of the theory, $\sigma$ is a parameter that numerates the points of the string, $\kappa$ is a dimensional parameter of the theory usually identified with the string tension, $[\kappa] = \text{GeV}^2$ (the system of units where $\hbar = c = 1$ is used).

By the standard procedure of equating the variation of the action (\ref{eq:NGaction}) to zero, we can obtain the equations of motion and boundary conditions. However, in their original form, the equations of motion are too complex to attempt to solve them directly. Fortunately, the Nambu-Goto action (\ref{eq:NGaction}) is re-parameterization invariant. This means that if the parameters $\tau$, $\sigma$ are replaced by arbitrary functions of these parameters, the value of the action (\ref{eq:NGaction}) will not change. This is understandable because, from a geometric point of view, the Nambu-Goto action is the invariant area of the world surface of the string taken with a dimensional coefficient. The area can be calculated in an arbitrary coordinate system. So, we will assume that such a parameterization of $\tau$, $\sigma$ is chosen, that it significantly simplifies the expression (\ref{eq:NGaction}) (this is called gauge selection). Usually, the orthonormal gauge (ONG) is used:
\begin{equation}
    \label{eq:ONG}
    \dot{x}^2+{x^{\prime}}^2=0,\quad \dot{x}x^{\prime}=0.
\end{equation}
From a geometric point of view, conditions (\ref{eq:ONG}) mean that an isometric (or conformal) system of curvilinear coordinates $\tau$, $\sigma$ is chosen on the world surface of the string.

In the orthonormal gauge one can obtain the equations of motion:
\begin{equation*}
    \ddot{x}_{\mu}-x^{\prime\prime}_{\mu}=0
\end{equation*}
and the boundary conditions of a string with free ends:
\begin{equation*}
    x^{\prime}_{\mu}(\tau,0)=x^{\prime}_{\mu}(\tau,\pi)=0.
\end{equation*}

Let us now assume that the functions $x_{\mu}(0,\sigma) = \rho_{\mu}(\sigma)$, $\dot{x}_{\mu}(0,\sigma) = v_{\mu}(\sigma)$, $\sigma \in [0, \pi]$ are known (they are called initial conditions or initial data of the problem). Then one obtains the following statement of the problem for the coordinates of a finite open relativistic string with free ends (the Cauchy boundary-value problem):
\begin{equation}
    \label{eq:CauchyProb}
    \begin{aligned}
         \ddot{x}_{\mu}-x^{\prime\prime}_{\mu} &= 0,\quad \sigma \in\left[0, \pi\right], \quad \tau>0, \quad \mu=0,~\ldots,~3;\\
         x^{\prime}_{\mu}(\tau,0) & =x^{\prime}_{\mu}(\tau,\pi) = 0;\\
         x_{\mu}(0,\sigma) & = \rho_{\mu}(\sigma), \quad \dot{x}_{\mu}(0,\sigma) = v_{\mu}(\sigma).
    \end{aligned}
\end{equation}

Since the solution of problem (\ref{eq:CauchyProb}) is well known in the subject of equations of mathematical physics, the main task in developing the model is to select the initial condition functions $\rho_{\mu}(\sigma)$ and $v_{\mu}(\sigma)$ that determine the coordinates and velocity of the string points at the initial moment in time. Choosing the specific type of initial condition functions is often called gauge selection, as this action finally fixes the relationship between the parameters $\tau$, $\sigma$ and the space-time coordinates.

The solution to the problem (\ref{eq:CauchyProb}) has the following form (see Appendix \ref{sec:appA} for more details):
\begin{equation}
    \label{eq:solution_std}
    x_{\mu}(\tau,\sigma) = Q_{\mu} + \frac{P_{\mu}\tau}{\kappa \pi} + \frac{i}{\sqrt{\kappa \pi}} \sum_{n \ne 0} e^{-in\tau}\frac{\alpha_{n\mu}}{n}\cos(n\sigma),\quad \mu = 0,~\ldots,~3,
\end{equation}
where
\begin{equation*}
    P_{\mu} \equiv \kappa \int_{0}^{\pi} \dot{x}_{\mu}(0,\sigma) d\sigma = \kappa \int_{0}^{\pi} v_{\mu}(\sigma) d\sigma
\end{equation*}
is the string total momentum 4-vector and
\begin{equation*}
    Q_{\mu} \equiv \frac{1}{\pi} \int_{0}^{\pi} x_{\mu}(0,\sigma) d\sigma = \frac{1}{\pi} \int_{0}^{\pi} \rho_{\mu}(\sigma) d\sigma
\end{equation*}
are the coordinates of the center of mass of the string at $\tau=0$.
The Fourier amplitudes $\alpha_{n\mu}$ are calculated in the following way:
\begin{equation}
    \label{eq:famp_std}
    \begin{aligned}
        \alpha_{n\mu} & = \sqrt{\frac{\kappa}{\pi}} \int_{0}^{\pi}\left[ v_{\mu}(\sigma) - in \rho_{\mu}(\sigma) \right]  \cos(n\sigma) d\sigma, \quad n \ne 0,\\
        \alpha_{0\mu} & =\frac{P_{\mu}}{\sqrt{\kappa \pi}}. 
    \end{aligned}
\end{equation}
The definition of $\alpha_{0\mu}$ is introduced for uniformity reasons.

A very important note to make is that the problem (\ref{eq:CauchyProb}) which produces Eq. (\ref{eq:solution_std}) does not define the possible motion of the string itself. Such equations of motion appear only if the ONG (\ref{eq:ONG}) is imposed. That means that the function (\ref{eq:solution_std}) can indeed describe the motion of the relativistic string only when it satisfies the orthonormal gauge. The conditions that arise after the substitution of formula (\ref{eq:solution_std}) in Eq. (\ref{eq:ONG}) can be expressed in terms of Fourier amplitudes:
\begin{equation}
    \label{eq:virasoro_cond}
    \sum_{m=-\infty}^{+\infty} \alpha_{n-m} \alpha_{m} = 0,\quad n = 0,~ \pm1,~ \pm2,~ \ldots ~
\end{equation}
The relations (\ref{eq:virasoro_cond}) are called Virasoro conditions for the relativistic string and were first considered within the framework of the dual-resonance models \cite{virasoro_1, virasoro_2}.

These conditions impose restrictions on the functions of the initial data $\rho_{\mu}(\sigma)$, $v_{\mu}(\sigma)$, meaning that their choice is not arbitrary. As will be shown in the next Section, finding the correct functional form of the initial conditions turns out to be a challenge. The first problem to be solved is developing a method that allows one to consistently and correctly specify the initial conditions of an arbitrarily moving massive relativistic string with free ends.

\section{On the problem of defining the initial conditions for massive relativistic string}
\label{sec:ICprob}
In all existing hadronization models that use the Nambu-Goto string approach the following ansatz \cite{CltchII_1, CltchII_2, CltchII_3} originally proposed by D.A. Morris for the Caltech-II model is used:
\begin{equation}
    \label{eq:MorrisAnzatz}
    \begin{aligned}
         \rho_{\mu}(\sigma) & \equiv 0,\quad v_{\mu}(\sigma) = v_{k\mu},\\
         \sigma_{k}  \le \sigma \le \sigma_{k+1},\quad k & = 1,~\ldots,~N,\quad \mu=0, 
~\ldots,~3,\\
    \end{aligned}
\end{equation}
where $N$ is the number of the segments of the string; the segmentation is usually introduced in string models of hadron production to take into account the hard gluons that are mapped on the string between the end-point partons. This means that the string is considered to be an initially point-like object with piece-wise constant function of velocity. Apart from the string being defined as spinless, the problem is that this approach turns out to be suitable only for massless strings, as will be shown further.

For demonstration purposes, a more general case will be considered here. Let the string initial conditions be defined as $\sigma$-wise linear functions (now, for simplicity, I will consider only one-segment strings; the generalization for the case of the string with gluons will be discussed later): 
\begin{equation}
    \label{eq:gen_initcond}
    v_{\mu}(\sigma) = A_{\mu} \sigma + B_{\mu}, \quad \rho_{\mu} (\sigma) = C_{\mu} (\sigma) + D_{\mu}, \quad \mu = 0,~\ldots,~3.
\end{equation}
Here $A_{\mu}$, $B_{\mu}$, $C_{\mu}$ and $D_{\mu}$ are the unknown numeric parameters that we will try to constrain by imposing the Virasoro conditions (\ref{eq:virasoro_cond}) on functions (\ref{eq:gen_initcond}). The functions (\ref{eq:gen_initcond}) are the first-order generalization of the Morris ansatz (\ref{eq:MorrisAnzatz}). As there are more free parameters in this case, it should be easier to satisfy the ONG gauge.

Calculating Fourier amplitudes using Eq. (\ref{eq:famp_std}) gives:
\begin{equation*}
    \alpha_{n\mu} = \sqrt{\frac{\kappa}{\pi}} \frac{{(-1)}^n-1}{n^2}\left( A_{\mu} - inC_{\mu} \right),\quad n \ne 0.
\end{equation*}
When substituting the Fourier amplitudes in Virasoro conditions, several cases arise. When the number $n$ is odd, then:
\begin{equation*}
    \sum_{m=-\infty}^{+\infty} \alpha_{n-m} \alpha_{m} = \frac{4 \kappa}{\pi} \left[ \sum_{k=-\infty}^{+\infty} \frac{A^2-inAC}{{\left( 2k+1 \right)}^2{\left( n-2k-1 \right)}^2} - \sum_{k=-\infty}^{+\infty} \frac{C^2}{{\left( 2k+1 \right)}^2\left( n-2k-1 \right)} \right] = 0.
\end{equation*}
This yields the following equations for the free parameters:
\begin{equation}
    \label{eq:gic_feq}
    A^2 = 0,\quad C^2 = 0, \quad AC = 0.
\end{equation}

If $n \ne 0$ is even, all the regular terms in the series are zero, but the ones that include $\alpha_{0\mu}$ remain:
\begin{equation*}
    -\frac{4}{n^2}\frac{AP-inCP}{\pi} = 0.
\end{equation*}
This produces the second set of equations:
\begin{equation}
    \label{eq:gic_seq}
    AP = BP,\quad PC = 0.
\end{equation}

When $n=0$, then:
\begin{equation*}
    \sum_{m=-\infty}^{+\infty} \alpha_{-m} \alpha_{m} = \frac{P^2}{\kappa \pi} + \frac{4 \kappa}{\pi} \left[ 2A^2 \sum_{k=-\infty}^{+\infty} \frac{1}{{\left( 2k+1 \right)}^4} + C^2 \sum_{k=-\infty}^{+\infty} \frac{1}{{\left( 2k+1 \right)}^2} \right] = 0.
\end{equation*}
This case produces the equation:
\begin{equation}
    \label{eq:gic_teq}
    \frac{P^2}{\kappa \pi} + \kappa \pi \left( \frac{A^2}{16} + C^2 \right) = 0.
\end{equation}
When Eqs. (\ref{eq:gic_feq}), (\ref{eq:gic_seq}) and (\ref{eq:gic_teq}) are considered together, the only way to make them solvable is to have the string with $P^2=0$, i.e., with zero mass.

Let us highlight the significance of the problem that we faced. The relativistic string as defined in Caltech-II and other similar string models simply does not exist in the Nambu-Goto theory! The used definition of the string is forbidden by the gauge imposed to obtain the equations of motion.

Let us consider the restrictions imposed by the Virasoro conditions (\ref{eq:virasoro_cond}) on the functions of the initial data in the general case. We write the Fourier amplitudes (\ref{eq:famp_std}) in the following form:
\begin{equation*}
    \alpha_{n\mu} \equiv f_{n\mu} - in g_{n\mu}, \quad n \ne 0,
\end{equation*}
where
\begin{equation*}
    f_{n\mu} = \sqrt{\frac{\kappa}{\pi}} \int_{0}^{\pi}d\sigma \cos(n\sigma) v_{\mu}(\sigma), \quad g_{n\mu} = \sqrt{\frac{\kappa}{\pi}} \int_{0}^{\pi}d\sigma \cos(n\sigma) \rho_{\mu}(\sigma).
\end{equation*}
Let us now substitute the coefficients written in this way into the Virasoro conditions (\ref{eq:virasoro_cond}):
\begin{equation*}
    \left\{
    \begin{aligned}
        \sum_{m \ne 0, m \ne n} \left( f_{n-m}f_{m}-m(n-m)g_{n-m}g_{m} \right) + \frac{2f_{n}P}{\sqrt{\kappa \pi}} & = 0,\quad n \ne 0\\
        \sum_{m \ne 0, m \ne n} \left( mf_{n-m}g_{m} + (n-m) f_{m}g_{n-m} \right) + \frac{2ng_{n}P}{\sqrt{\kappa \pi}} & = 0, \quad n \ne 0\\
        \sum_{m \ne 0} \left( f_{-m}f_{m}+m^2g_{-m}g_{m} \right) +  \frac{P^2}{\kappa \pi} & = 0\\
        \sum_{m \ne 0} m\left( f_{-m}g_{m} - f_{m}g_{-m} \right) & = 0.\\
    \end{aligned}
    \right.
\end{equation*}

The types of functions $\rho_{\mu}(\sigma)$, $v_{\mu}(\sigma)$ that satisfy these expressions can be divided into three categories. The first category includes functions for which all scalar products $f_{n-m}f_m$, $g_{n-m} g_m$, $f_{n-m} g_m$ are equal to zero. This is the ``strongest''  fulfilling of the Virasoro conditions, but such functions lead to too strong restrictions on the string parameters, e.g., to the requirement that the string mass must be zero. The second category consists of the functions $\rho_{\mu}(\sigma)$, $v_{\mu}(\sigma)$ for which each individual term in the sums is equal to zero. This requires that the coefficients $m$, $(n-m)$, $m(n-m)$ explicitly disappear from the expressions. So, due to the arbitrariness of the numbers $n$, $m$ it would not be necessary to make each of the scalar products equal to zero separately. It is difficult to select such a special type of functions, because it is necessary that the following integral, e.g., was representable as
\begin{equation*}
    g_{n\mu} = \sqrt{\frac{\kappa}{\pi}} \int_{0}^{\pi}d\sigma \cos(n\sigma)\rho_{\mu}(\sigma) \equiv \frac{g_{\mu}}{n},
\end{equation*}
which requires a very inventive selection of functions $\rho_{\mu}(\sigma)$, $v_{\mu}(\sigma)$.

The third category of initial data functions includes those $\rho_{\mu}(\sigma)$, $v_{\mu}(\sigma)$ for which the Virasoro conditions are satisfied in the ``weakest'' sense, i.e. only the infinite sum itself is equal to zero. Even if one considers the third category of initial data functions, the non-zero mass may still be forbidden (as in the case of the initial conditions in the form (\ref{eq:gen_initcond})).

Thus, it is clear that the selection of initial data functions satisfying the Virasoro conditions and allowing the string to have mass is not a trivial task. It can be shown that polynomials of any degree, fractional rational expressions, functions containing sine, combinations of Dirac delta functions, and power series are not suitable for this purpose.

\section{The FOEE method to define the initial conditions}
\label{sec:FOEEmethod}
In this paper, to solve the problem of specifying the initial data for the string that satisfies the Virasoro conditions, I propose a new method based on the finite expansion of the functions $\rho_{\mu}(\sigma)$, $v_{\mu}(\sigma)$ in eigenfunctions of the Sturm-Liouville problem (see Appendix \ref{sec:appA}) arising when solving the Cauchy problem for the motion of a string (\textbf{F}inite-\textbf{O}rder \textbf{E}igenfunction \textbf{E}xpansion, \textbf{FOEE}).

The essence of the method is as follows. Let the functions of the initial data be represented as series:
\begin{equation}
    \label{eq:foee_ic_base}
    \begin{aligned}
        v_{\mu}(\sigma)  = a_{0\mu}u_{0}(\sigma) + \sum_{k \ne 0} a_{k\mu}u_{k}(\sigma), \quad \rho_{\mu}(\sigma)  = b_{0\mu}u_{0}(\sigma) + \sum_{k \ne 0} b_{k\mu}u_{k}(\sigma).\\
    \end{aligned}
\end{equation}
Here, $u_k(\sigma)$, $\sigma \in [0, \pi]$ is the $k$-th eigenfunction of the Sturm problem. In the case of a free string (see Appendix \ref{sec:appA}) $u_k(\sigma) = \cos(k\sigma)$, $u_0(\sigma) \equiv 1$. Then the functions of the initial data (\ref{eq:foee_ic_base}) take the form:
\begin{equation}
    \label{eq:foee_ic_fs}
    \begin{aligned}
        v_{\mu}(\sigma) = a_{0\mu} + \sum_{k \ne 0} a_{k\mu} \cos(k\sigma), \quad \rho_{\mu}(\sigma) = b_{0\mu} + \sum_{k \ne 0} b_{k\mu} \cos(k\sigma).\\
    \end{aligned}
\end{equation}
Let us consider, what the Fourier amplitudes $\alpha_{n\mu}$ are equal to in this case:
\begin{equation*}
    \begin{aligned}
        \alpha_{n\mu} &= \sqrt{\frac{\kappa}{\pi}} \int_{0}^{\pi}d\sigma \cos(n\sigma) \left[ a_{0\mu} + \sum_{k \ne 0} a_{k\mu} \cos(k\sigma) - in \left( b_{0\mu} + \sum_{k \ne 0} b_{k\mu} \cos(k\sigma) \right) \right]\\
        &= \sqrt{\frac{\kappa}{\pi}} \frac{\pi}{2} \sum_{k \ne 0} \left[ a_{k\mu} \delta_{nk} - in b_{k\mu} \delta_{nk} \right] = \frac{\sqrt{\kappa \pi}}{2} \left( a_{n\mu} - in b_{n\mu} \right),
    \end{aligned}
\end{equation*}
where $\delta_{nk}$ is the Kronecker delta. The orthogonality property of the eigenfunctions (\ref{eq:Aef_ort}) was used in the derivation process.

The Virasoro conditions are now written as follows:
\begin{equation}
    \label{eq:virasoro_foee_gen}
    \left\{
    \begin{aligned}
        \sum_{m \ne 0, m \ne n} \left( a_{n-m}a_{m}-m(n-m)b_{n-m}b_{m} \right) +  \frac{4}{\kappa \pi} Pa_n & = 0,\quad n \ne 0\\
        \sum_{m \ne 0, m \ne n} \left( ma_{n-m}b_{m} + (n-m) a_{m}b_{n-m} \right) + \frac{4n}{\kappa \pi} Pb_n & = 0, \quad n \ne 0\\
        \sum_{m \ne 0} \left( a_{-m}a_{m}+m^2b_{-m}b_{m} \right) +  \frac{2P^2}{{(\kappa \pi)}^2} & = 0\\
        \sum_{m \ne 0} m\left( a_{-m}b_{m} - a_{m}b_{-m} \right) & = 0.\\
    \end{aligned}
    \right.
\end{equation}

These equalities are satisfied for all values $n \in \mathbb{Z}$ . If the functions in (\ref{eq:foee_ic_fs}) are represented by infinite series, this leads to an infinite number of equations of the form (\ref{eq:virasoro_foee_gen}). Assuming that all coefficients $a_{n\mu}$, $b_{n\mu}$, starting from some $|n|>N$, are equal to zero, one obtains finite series in expansions (\ref{eq:foee_ic_fs}) and a finite number of equations for the remaining coefficients in (\ref{eq:virasoro_foee_gen}). This is the essence of the FOEE method: now the system has a finite number of equations, which means that there is no need to consider the sum values in the (\ref{eq:virasoro_foee_gen}) for arbitrary numbers $n$, $m$. The value of $N$, which marks the largest absolute value of the order of the non-zero Fourier amplitude hereinafter will be referred to as the ``order'' of the FOEE system (expansion).

In total, there are $2(4N+1)$ equations and $8(2N+1)$ unknown coefficients. One should add conservation laws to accompany equations (\ref{eq:virasoro_foee_gen}):
\begin{equation}
    \label{eq:consP_base}
    \begin{aligned}
        P_{\mu} \equiv \kappa \int_{0}^{\pi} d\sigma v_{\mu}(\sigma) = \kappa \int_{0}^{\pi} d\sigma \left( a_{0\mu} + \sum_{k \ne 0} a_{k\mu} \cos(k\sigma) \right) = \kappa \pi a_{0\mu},
    \end{aligned}
\end{equation}
\begin{equation}
    \label{eq:consM_base}
    \begin{aligned}
        \mathcal{M}_{\mu \nu} & \equiv \kappa \int_{0}^{\pi} d\sigma \left[ \rho_{\mu}(\sigma) v_{\nu}(\sigma) - \rho_{\nu}(\sigma) v_{\mu}(\sigma) \right] \\
        &= \kappa \int_{0}^{\pi} d\sigma \left[ \left( b_{0\mu} + \sum_{k \ne 0} b_{k\mu} \cos(k\sigma) \right) \left( a_{0\nu} + \sum_{l \ne 0} a_{l\nu} \cos(l\sigma) \right) \right.\\
        & \quad \quad \quad \quad ~~\left.- \left( b_{0\nu} + \sum_{k \ne 0} b_{k\nu} \cos(k\sigma) \right) \left( a_{0\mu} + \sum_{l \ne 0} a_{l\mu} \cos(l\sigma) \right) \right]\\
        &= \kappa \pi \left( b_{0\mu}a_{0\nu} - b_{0\nu}a_{0\mu} \right) + \kappa \sum_{k \ne 0} \sum_{l \ne 0} \left( b_{k\mu}a_{l\nu} - b_{k\nu}a_{l\mu} \right) \int_{0}^{\pi}d\sigma \cos(k\sigma) \cos(l\sigma)\\
        &= b_{0\mu}P_{\nu} - b_{0\nu}P_{\mu} + \frac{\kappa \pi}{2} \sum_{k \ne 0}  \left( b_{k\mu}a_{k\nu} - b_{k\nu}a_{k\mu} \right).
    \end{aligned}
\end{equation}
In deriving these relations, the orthogonality property (\ref{eq:Aef_ort}) of the eigenfunctions was used again. The first equation corresponds to the conservation of the total 4-momentum of the system $P_{\mu}$, the second one corresponds to the conservation of the total angular momentum tensor $\mathcal{M}_{\mu \nu}$.

Taking into account the symmetry properties of the total angular momentum tensor $\mathcal{M}_{\mu \nu}$, one obtains that the conservation laws add another $4+6=10$ equations for the coefficients $a_{n\mu}$, $b_{n\mu}$, where $n = 0,~\pm1,~\ldots,~\pm N$. Thus, we have a system of $8N+12$ equations and $16N+8$ unknown parameters. Obviously, for $N>0$ the system is not overloaded with equations, and the existence of a solution is allowed (which possibly requires additional assumptions).

Note that in this method, it is assumed that when the string is formed, the coordinates of the partons are unknown. Firstly, it is indeed unclear when, after the production of partons, the preconfinement, i.e. the formation of strings, occurs. The assumptions about this could be made based on the study of the thermal dependence of the string tension coefficient $\kappa$ \cite{Hunt_Smith_2020}. In the absence of an alternative, it is reasonable to assume that the spatial configuration of the system is determined precisely from the relativistic string model, which currently provides the most realistic description of the dynamics of a bound system of partons. In addition, if we assume that the coordinates of the quarks at the ends of the string are known, this will add another $8$ equations and increase the minimum order of the expansion (\ref{eq:foee_ic_fs}) from $N=1$ to $N=2$, which will significantly complicate solving the system.

Because of its non-linear nature, solving the complete system of equations composed of (\ref{eq:virasoro_foee_gen}) and (\ref{eq:consP_base}), (\ref{eq:consM_base}) turns out to be quite a complicated task even for the second order of the expansion, i.e. for two non-zero Fourier amplitudes: $\alpha_{\pm1 \mu}$, $\alpha_{\pm2 \mu}$.

\section{The FOEE(1)-string}
\label{sec:FOEE_1}
This Section is dedicated to the simplest case of the FOEE system (\ref{eq:virasoro_foee_gen}), when only one order of Fourier amplitudes is non-zero. I refer to the relativistic string, that is defined in this way, as FOEE(1)-string. In this case, the functions of initial conditions for the Cauchy problem (\ref{eq:CauchyProb}) are written as:
\begin{equation}
    \label{eq:foee1_ic_base}
    \begin{aligned}
        v_{\mu}(\sigma) = a_{\mu} + b_{\mu} \cos(\sigma), \quad
        \rho_{\mu}(\sigma) = c_{\mu} + d_{\mu} \cos(\sigma).
    \end{aligned}
\end{equation}

Of course, defining the string oscillation by a single eigenharmonic is a significant simplification. However, it is not yet known whether the cases of higher order are solvable; at least no single solution was found yet even for the second order (not even partial one). It might be true that the FOEE(1)-string is a formidable approximation to model the general string motion. In any case, it can still be considered as a better approximation than point-like strings without rotation.

The formulae for the Fourier amplitudes are the following (including the zero-order term):
\begin{equation}
    \label{eq:foee1_famp_base}
    \begin{aligned}
        \alpha_{0\mu} = \frac{P_{\mu}}{\sqrt{\kappa \pi}}, \quad
        \alpha_{1\mu} = \frac{\sqrt{\kappa \pi}}{2} \left( b_{\mu} - id_{\mu} \right), \quad
        \alpha_{-1\mu} = \frac{\sqrt{\kappa \pi}}{2} \left( b_{\mu} + id_{\mu} \right).\\
    \end{aligned}
\end{equation}

To construct the Virasoro system, the values $n=-2,~\ldots,~2$ should be considered in (\ref{eq:virasoro_foee_gen}). After excluding the repeating expressions one obtains the system of the Virasoro conditions for the FOEE(1)-string:
\begin{equation}
    \label{eq:foee1_virasoro}
    \left\{
    \begin{aligned}
        b^2 - d^2 &= 0\\
        bd = bP = dP &= 0\\
        b^2 + d^2 + \frac{2P^2}{(\kappa \pi)^2} &= 0.
    \end{aligned}
    \right.
\end{equation}

The laws of energy-momentum and angular momentum conservation according to Eqs. (\ref{eq:consP_base}), (\ref{eq:consM_base}) take the form:
\begin{equation}
    \label{eq:foee1_consP}
    \begin{aligned}
        a_{\mu} = \frac{P_{\mu}}{\kappa \pi},
    \end{aligned}
\end{equation}
\begin{equation}
    \label{eq:foee1_consM}
    \begin{aligned}
        \mathcal{M}_{\mu \nu} = c_{\mu}P_{\nu} - c_{\nu}P_{\mu} + \frac{\kappa \pi}{2} \left( d_{\mu}b_{\nu} - d_{\nu}b_{\mu} \right).
    \end{aligned}
\end{equation}

It is quite evident from system (\ref{eq:foee1_virasoro}) that the oscillation has to be present for both the initial velocity function $v_{\mu}(\sigma)$ and the initial coordinates of the string $\rho_{\mu}(\sigma)$, otherwise the string must be considered massless. The relation (\ref{eq:foee1_consM}) is of particular importance, as it shows that even for the string without rotation (as is usually accepted) the certain relation between coefficients $b_{\mu}$ and $d_{\mu}$, $\mu=0,~\ldots,~3$, exists nevertheless and cannot be neglected for the massive string.

\subsection{The FOEE(1)-string in the rest frame}
\label{sec:restfr}
Since it is difficult to find a solution to the system of (\ref{eq:foee1_virasoro}) and (\ref{eq:foee1_consM}) in the general case due to its nonlinearity, let us first consider the simplest case. Let a relativistic string of mass $M$ be defined in the system of its center of mass: $P_0 \equiv M$, $P_i \equiv 0$, $i=1,~2,~3$. We leave one of the independent components of the angular momentum tensor non-zero in order to consider a string with rotation (the equality of the remaining components to zero can always be achieved by rotating the coordinate system): $M_{13} \ne 0$, $M_{\mu\nu} \equiv 0$, $\mu \ne 1$, $\nu \ne 3$, $\mu \le \nu$. Also note, this means that the coordinate system is placed in a way that its origin coincides with the center of mass of the string. The assumption that the rotation of the string occurs in a plane comes from the fact that the string is defined initially by two point-like partons. Then we obtain the following form of the FOEE system:
\begin{equation}
    \label{eq:foee1_cm_syst}
    \left\{
    \begin{aligned}
        b^2 - d^2 = bd = b_0M = d_0M &= 0\\
        b^2 - \frac{M^2}{(\kappa \pi)^2} &= 0\\
        - c_{1}M + \frac{\kappa \pi}{2} \left( d_{0}b_{1} - d_{1}b_{0} \right) &= 0\\
        - c_{2}M + \frac{\kappa \pi}{2} \left( d_{0}b_{2} - d_{2}b_{0} \right) &= 0\\
        - c_{3}M + \frac{\kappa \pi}{2} \left( d_{0}b_{3} - d_{3}b_{0} \right) &= 0\\
        d_{1}b_{2} - d_{2}b_{1} = d_{2}b_{3} - d_{3}b_{2} &= 0\\
        d_{1}b_{3} - d_{3}b_{1} - \frac{2 \mathcal{M}_{1 3}}{\kappa \pi} &= 0.
    \end{aligned}
    \right.
\end{equation}

One can immediately see that $b_0=d_0=0$. Taking this into account, the first three equations from the conservation of angular momentum tensor give $c_1=c_2=c_3=0$. The value of $c_0$ (it corresponds to the value of time $t$ at $\tau=0$) remains arbitrary and, therefore, let us also choose it as zero. The system
\begin{equation*}
    \left\{
    \begin{aligned}
        d_{1}b_{2} - d_{2}b_{1} =d_{2}b_{3} - d_{3}b_{2} &= 0\\
        d_{1}b_{3} - d_{3}b_{1} - \frac{2 \mathcal{M}_{1 3}}{\kappa \pi} &= 0
    \end{aligned}
    \right.
\end{equation*}
does not lead to a contradiction only if $b_2=d_2=0$. Then only the following equations remain:
\begin{equation}
    \label{eq:temp_syst}
    \left\{
    \begin{aligned}
        b_1d_1 + b_3d_3 &= 0\\
        b_1^2 + b_3^2 - d_1^2 - d_3^2 &= 0\\
        b_1^2 + b_3^2 - \frac{M^2}{(\kappa \pi)^2} &= 0\\
        d_1b_3 - d_3b_1 - \frac{2 \mathcal{M}_{13}}{\kappa \pi} &= 0.
    \end{aligned}
    \right.
\end{equation}
The expression for $d_3$ can be obtained using the first equation and then substituted into the fourth. This will produce:
\begin{equation*}
    \begin{aligned}
        d_1 = \frac{2 \kappa \pi \mathcal{M}_{13}}{M^2} b_3, \quad d_3 = - \frac{2 \kappa \pi \mathcal{M}_{13}}{M^2} b_1.
    \end{aligned}
\end{equation*}
If we now substitute $d_1$ and $d_3$ into the second equation in (\ref{eq:temp_syst}), we get
\begin{equation*}
    \begin{aligned}
        \left( b_1^2 + b_3^2 \right) \left( 1 - \left( \frac{2 \kappa \pi \mathcal{M}_{13}}{M^2} \right)^2 \right) = 0.
    \end{aligned}
\end{equation*}
Since the sum $b_1^2 + b_3^2$ can equal to zero only in the case of a massless string (according to the third equation in (\ref{eq:temp_syst})), the only choice left is to require the fixed relation between the mass of the string and its spin:
\begin{equation}
    \label{eq:spin_mass1}
    2 \kappa \pi |\mathcal{M}_{13}| = M^2.
\end{equation}
To solve the system, an additional equation is required now, as Eq. (\ref{eq:spin_mass1}) does not constrain the values of $b_{1,3}$ and $d_{1,3}$. Let $b_1=0$, then:
\begin{equation}
    \label{eq:foee1_cm_coef_solved}
    b_1 = d_3 = 0, \quad b_3 = \xi \frac{M}{\kappa \pi}, \quad d_1 = \frac{M}{\kappa \pi}
\end{equation}
where the signature of the string rotation $\xi$ is introduced:
\begin{equation}
    \label{eq:xi}
    \xi = \text{sign}(\mathcal{M}_{13}).
\end{equation}

The system is now completely solved. The initial conditions for the FOEE(1)-string are defined by the formulae:
\begin{equation}
    \label{eq:foee1_ic_cm}
    \begin{aligned}
        v_{\mu}(\sigma) = (\kappa \pi)^{-1}M \left[ \delta_{0\mu} + \xi \delta_{3\mu} \cos(\sigma) \right], \quad
        \rho_{\mu}(\sigma) = (\kappa \pi)^{-1}M \delta_{1\mu}\cos(\sigma), 
    \end{aligned}
\end{equation}
where $\delta_{\mu \nu}$ is the Kronecker delta.

The Fourier amplitudes for the FOEE(1)-string are calculated as:
\begin{equation}
    \label{eq:foee1_famp_cm}
    \begin{aligned}
        \alpha_{n\mu} &= \sqrt{\frac{\kappa}{\pi}} \int_{0}^{\pi} d\sigma \frac{M}{\kappa \pi} \left[ \xi \delta_{3\mu} - in \delta_{1\mu}  \right] \cos(n\sigma) \cos(\sigma)\\
        &= \frac{M}{2\sqrt{\kappa \pi}} \left[ \xi \delta_{3\mu} - in \delta_{1\mu}  \right] \delta_{1|n|},\quad n \ne 0.
    \end{aligned}
\end{equation}
As expected, the only amplitudes remaining non-zero are $\alpha_{\pm1}$. One can easily check that Fourier amplitudes satisfy the Virasoro conditions:
\begin{equation*}
    (\alpha_{\pm1\mu})^2 = \alpha_{\pm1}\alpha_{0} = 2\alpha_{-1}\alpha_{1} + (\alpha_{0\mu})^2 = 0.
\end{equation*}

The tensor of the angular momentum defined by the inner rotation of the string (spin tensor) can be written in the following form:
\begin{equation}
    \label{eq:spin_tensor}
    S_{\mu \nu} = - \frac{i}{2} \sum_{n \ne 0} (\alpha_{-n\mu}\alpha_{n\nu} - \alpha_{-n\nu}\alpha_{n\mu}) \equiv \mathcal{M}_{13}
\end{equation}
and matches the total angular momentum $\mathcal{M}_{\mu \nu}$ of the string defined in the center-of-mass frame.

Let us also recall that in the Nambu-Goto string theory the following inequality is derived, which is true for any string (defined in the static gauge, i.e. when $x_0 (\tau,\sigma) \sim \tau)$ \cite{barbashov_nesterenko}:
\begin{equation}
    \label{eq:spin_limit}
    J \le \frac{M^2}{2 \kappa \pi},
\end{equation}
where $J$ is the classical value of the spin of the string. The universal expression for the spin of the string has the form \cite{barbashov_nesterenko}:
\begin{equation}
    \label{eq:spin_string}
    J^2 = \frac{1}{2} \left( S_{\mu \nu}S^{\mu \nu} - \frac{2}{M^2} P_{\nu}S^{\nu \rho}P^{\sigma}S_{\sigma \rho} \right).
\end{equation}
Since there is no motion of the string as a whole in the rest frame, the spin $J$ is a total angular momentum of the string: $J = M^2 / (2 \kappa \pi)$. Thus, the FOEE(1)-string defined by Eq. (\ref{eq:foee1_ic_cm}) turns out to be an extreme case of (\ref{eq:spin_limit}).

The expression for the string coordinates $x_{\mu}(\tau,\sigma)$ can be written in the following way:
\begin{equation*}
    x_{\mu}(\tau,\sigma) = \frac{M}{\kappa \pi} \left[ \delta_{0\mu} \tau + \frac{i}{2} \left( [\xi\delta_{3\mu} - i\delta_{1\mu}] e^{-i\tau} - [\xi\delta_{3\mu} + i\delta_{1\mu}] e^{i\tau} \right) \cos(\sigma) \right].
\end{equation*}
Using the Euler's formula one can simplify the expression to obtain the final form:
\begin{equation}
    \label{eq:foee1_cm_solution}
    x_{\mu}(\tau,\sigma) = (\kappa \pi)^{-1}M \left( \delta_{0\mu} \tau + \left[ \xi \delta_{3\mu}\sin(\tau) + \delta_{1\mu} \cos(\tau) \right] \cos(\sigma) \right).
\end{equation}
String velocity can be calculated as
\begin{equation}
    \label{eq:foee1_cm_xdot}
    \dot{x}_{\mu}(\tau,\sigma) =  (\kappa \pi)^{-1}M \left( \delta_{0\mu} + \left[ \xi \delta_{3\mu}\cos(\tau) - \delta_{1\mu} \sin(\tau) \right] \cos(\sigma) \right)
\end{equation}
and the derivative $x^{\prime}(\tau,\sigma)$
\begin{equation}
    \label{eq:foee1_cm_xprime}
    x^{\prime}_{\mu}(\tau,\sigma) = -(\kappa \pi)^{-1}M \left[ \xi \delta_{3\mu}\sin(\tau) +\delta_{1\mu} \cos(\tau) \right] \sin(\sigma).
\end{equation}

It is important to study one more relationship, the fulfillment of which is necessary for the correct definition of the string. These are the conditions imposed on the tangent vectors to the world sheet of the string. They are derived based on the requirement that the total velocity of the points of the string is less than the speed of light, which ensures the positive sign of the expression under the square root in the Nambu-Goto action:
\begin{equation*}
    (\dot{x}x^{\prime})^2 - \dot{x}^2x^{\prime2} > 0,
\end{equation*}
which in the orthonormal gauge (\ref{eq:ONG}) is equivalent of requiring
\begin{equation}
    \label{eq:tang_vec}
    \dot{x}^2 > 0, \quad x^{\prime2} < 0.
\end{equation}
Let us check the fulfillment of the conditions (\ref{eq:tang_vec}) for the initial data functions (\ref{eq:foee1_ic_cm}) and the derivatives of the coordinate function (\ref{eq:foee1_cm_xdot}), (\ref{eq:foee1_cm_xprime}):
\begin{equation*}
    \begin{aligned}
        \left( v_{\mu}(\sigma) \right)^2 &= \left( \dot{x}_{\mu}(\tau,\sigma) \right)^2 = \frac{M^2}{(\kappa \pi)^2} \sin(\sigma) \ge 0,\\
        \left( \rho_{\mu}^{\prime}(\sigma) \right)^2 &= \left( x_{\mu}^{\prime}(\tau,\sigma) \right)^2= -\frac{M^2}{(\kappa \pi)^2} \sin(\sigma) \le 0.
    \end{aligned}
\end{equation*}
This ends the proof that the functions (\ref{eq:foee1_ic_cm}) correctly define the initial conditions for the massive relativistic string in the rest frame.

\begin{figure}[htbp]
\centering
\includegraphics[width=.6\textwidth]{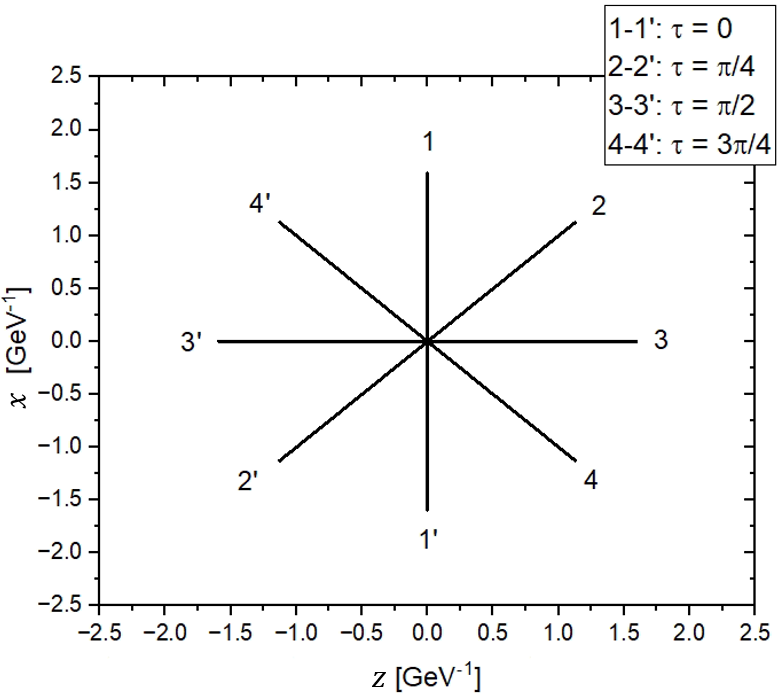}
\caption{The FOEE(1)-string in its rest frame rotates as a rigid rod.\label{fig:fig1}}
\end{figure}

Let us give an example of calculating the motion of a relativistic string with initial conditions specified according to (\ref{eq:foee1_ic_cm}). Let the string model a quark-antiquark system, and let its invariant mass be set to $1$ GeV. Using (\ref{eq:foee1_cm_solution}), one obtains that the string moves in its rest frame as shown in figure \ref{fig:fig1}. Note that according to the selected system of units, the spatial extension of the string is measured in $\text{GeV}^{-1}$.
One can see that the FOEE(1)-string defined in the center-of-mass system rotates as a rigid rod. At the time when $\tau=\pi$ the ends of the string change places compared to the initial moment in time. At $\tau=2\pi$, the string returns to its initial state.

\subsection{A string moving with momentum \texorpdfstring{$\boldsymbol{P}$}{\textbf{P}} relative to the rest frame}
\label{sec:foee_mov}
Now we need to generalize the simplest case of a string in the rest frame to the case of an arbitrarily moving string. First, consider a boost to a system where the string has momentum $\boldsymbol{P}$. Recall the Lorentz transformations for an ordinary particle. Let particle $1$ have energy $\omega_1^*$ and momentum $\boldsymbol{p}_1^*$ in the rest frame of particle $0$ with mass $m_0$. We need to determine the energy and momentum of the particle $1$ in a system in which the particle $0$ has momentum $\boldsymbol{p}_0$. The general Lorentz transformations can be written in a very convenient way as follows \cite{Kopylov}:
\begin{equation}
    \label{eq:lorentz_transform}
    \begin{aligned}
        \omega_1 = \frac{\omega_0 \omega_1^* + \boldsymbol{p}_0\boldsymbol{p}_1^*}{m_0}, \quad \boldsymbol{p}_1 = \boldsymbol{p}_1^* + \boldsymbol{p}_0 \frac{\omega_1 + \omega_1^*}{\omega_0 + m_0},
    \end{aligned}
\end{equation}
where $\omega_0 = \sqrt{m_0^2+\boldsymbol{p}_0^2}$. The convenience of formulae (\ref{eq:lorentz_transform}) lies in the ability to easily write down the transformation for any component separately. But calculations must be done in two stages (first the energy and then the momentum must be calculated). If, instead of particle $1$, one writes the transformations for particle $0$, expressions (\ref{eq:lorentz_transform}) turn into obvious identities.

In the case of a string, one has to work with distributed quantities that depend on the parameter $\sigma$. Let us perform the Lorentz transformation for each point of the string with coordinate $\sigma$. Instead of the energy $\omega_0$, momentum $\boldsymbol{p}_0$ and mass $m_0$ of the particle $0$, we can substitute the momentum $\boldsymbol{P}$, total energy $P_0=\sqrt{M^2 + \boldsymbol{P}^2}$ and mass of the string $M$. Instead of the energy $\omega_1^*$ of the particle $1$ in the rest frame of the particle $0$, we can take the distributed energy of the string $p_0^*(\sigma) \equiv \kappa v_0^*(\sigma)$, defined according to (\ref{eq:foee1_cm_xdot}). Finally, $\boldsymbol{p}_1^*$ should be replaced with $\boldsymbol{p}^*(\sigma) \equiv \kappa \boldsymbol{v}^*(\sigma)$. Then we obtain formulae for the Lorentz boost of a relativistic string with mass $M$ from the rest frame to the system where its momentum is equal to $\boldsymbol{P}$:
\begin{equation}
    \label{eq:lorentz_boost_string}
    \begin{aligned}
        p_0(\sigma) = \frac{P_0 p_0^*(\sigma) + \boldsymbol{P}\boldsymbol{p}^*(\sigma)}{M}, \quad\boldsymbol{p}(\sigma) = \boldsymbol{p}^*(\sigma) + \boldsymbol{P} \frac{p_0(\sigma) + p_0^*(\sigma)}{P_0 + M}.
    \end{aligned}
\end{equation}
Let us see how the transformations (\ref{eq:lorentz_boost_string}) will look for the initial conditions of the FOEE(1)-string (\ref{eq:foee1_ic_cm}). The distributed energy of the string in the new reference frame is
\begin{equation*}
    p_0(\sigma) = \frac{P_0 + \xi P_3\cos(\sigma)}{\pi},
\end{equation*}
the vector function of the string momentum per unit $\sigma$ is expressed as:
\begin{equation*}
    \begin{aligned}
        \boldsymbol{p}(\sigma) = \frac{\xi M}{\pi} \boldsymbol{e}_{z} \cos(\sigma) + \frac{\boldsymbol{P}}{\pi} \frac{P_0 + \xi P_3 \cos(\sigma) + M}{P_0 + M} =\frac{1}{\pi} \left( \boldsymbol{P} + \xi \left[\frac{P_3}{P_0+M}\boldsymbol{P} + M\boldsymbol{e}_z \right] \cos(\sigma) \right),
    \end{aligned}
\end{equation*}
where $\boldsymbol{e}_x = (1,~0,~0)$ is a unit vector in the direction of $X$-axis. The combining formula for the initial velocity function of the string is
\begin{equation}
    \label{eq:foee1_mov_v}
    v_{\mu}(\sigma) = \frac{P_{\mu}}{\kappa \pi} + \frac{\xi}{\kappa \pi} \left[ \frac{P_3 \left( P_{\mu} + M\delta_{0\mu} \right)}{P_0+M} + M\delta_{3\mu} \right] \cos(\sigma).
\end{equation}

The same procedure can by applied to the function of the string initial coordinates $\rho_{\mu}(\sigma)$:
\begin{equation*}
\begin{aligned}
        \rho_0(\sigma) &= \frac{P_0\rho_0^*(\sigma) + \boldsymbol{P}\boldsymbol{\rho}^*(\sigma)}{M} = \frac{P_1\cos(\sigma)}{\kappa\pi},\\
        \boldsymbol{\rho}(\sigma) &= \frac{M}{\pi} \boldsymbol{e}_{x} \cos(\sigma) + \frac{\boldsymbol{P}}{\kappa \pi} \frac{P_1 \cos(\sigma)}{P_0 + M}=\frac{1}{\pi} \left( \frac{P_1}{P_0+M}\boldsymbol{P} + M\boldsymbol{e}_x  \right) \cos(\sigma),
    \end{aligned}
\end{equation*}
which results in a formula:
\begin{equation}
    \label{eq:foee1_mov_rho}
    \rho_{\mu}(\sigma) = \frac{1}{\kappa \pi} \left[ \frac{P_1 \left( P_{\mu} + M\delta_{0\mu} \right)}{P_0+M} + M\delta_{1\mu} \right] \cos(\sigma).
\end{equation}
Thus, formulae (\ref{eq:foee1_mov_v}), (\ref{eq:foee1_mov_rho}) define the initial data for the massive relativistic string in the reference frame where it moves relative to its rest frame as a whole with momentum $\boldsymbol{P}$.

\subsection{The FOEE(1)-string defined in the arbitrary frame of reference}
\label{sec:foee_gen}
\begin{figure}[htbp]
\centering
\includegraphics[width=.5\textwidth]{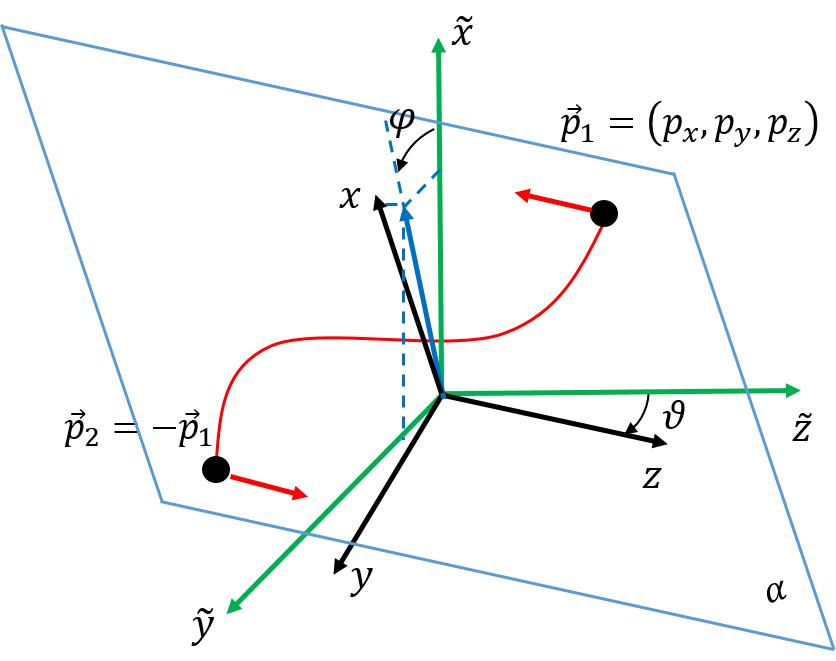}
\caption{The scheme of the rotation of the coordinate system axis. Rotation is performed in the center-of-mass system of the string.\label{fig:fig3}}
\end{figure}

In order to proceed to the consideration of the general case of string motion, it is necessary to add the transformation of the rotation of the coordinate system to the Lorentz boost. We start with two partons with arbitrary momenta defined in an arbitrary frame of reference. The next step would be to perform a Lorentz boost to their center-of-mass frame. The orientation of the coordinate system axis would remain the same. So in order to make the formulae (\ref{eq:foee1_ic_cm}) applicable, the coordinate system must be rotated in such a way that only one component of the string angular momentum tensor remains. This is illustrated in figure \ref{fig:fig3}. Let us denote the coordinate system in the center-of-mass frame with ``$\sim$'' sign. The coordinate system oriented to align its $Y$-axis with the direction of the string angular momentum will be marked with ``$*$'' symbol. Then the initial conditions (\ref{eq:foee1_ic_cm}) define the string in the ``$*$''-frame. For transition into the ``$\sim$''-frame, one must perform the counter-rotation. For this, it is sufficient to act on the spatial components of the initial conditions (\ref{eq:foee1_ic_cm}) by the rotation matrix
\begin{equation}
   \label{eq:rotation_mat}
   R(\theta,\varphi) =
    \begin{pmatrix}
    \cos\varphi & -\sin\varphi \cos\theta & \sin\varphi\sin\theta\\
    \sin\varphi & \cos\varphi\cos\theta & -\cos\varphi\sin\theta\\
    0 & \sin\theta & \cos\theta
    \end{pmatrix}
\end{equation}
to return to the coordinate system where momenta of two partons (in their center-of-mass frame) were initially defined (see figure \ref{fig:fig3}). Let one of the partons have momentum $\boldsymbol{p}_1=(p_x,~p_y,~p_z)$ in the ``$\sim$'' center-of-mass frame (the other one will obviously have the momentum $\boldsymbol{p}_2 = - \boldsymbol{p}_1$). Then the elements of the rotation matrix (\ref{eq:rotation_mat}) can be defined as:
\begin{equation}
    \label{eq:thetaphi}
    \begin{aligned}
        \cos\theta = \frac{p_z}{\sqrt{p_x^2 + p_y^2 + p_z^2}}, \quad \cos\varphi = \frac{p_x}{\sqrt{p_x^2 + p_y^2}}.
    \end{aligned}
\end{equation}
In the special case of $p_x^2 + p_y^2 = 0$ angles $\theta$, $\varphi$ should be equaled to zero. Note that the actual values of the momenta of two partons in ``$*$''-frame are not needed, as only the invariant mass of the string $M$ is required to fully define it in this frame.

We start with the initial conditions for the FOEE(1)-string defined in the center-of-mass frame with the coordinate system rotated in the way that only the rotation around $Y$-axis remains (the ``$*$''-frame):
\begin{equation}
    \label{eq:ic_cm_star}
    \begin{aligned}
        v^*_{\mu}(\sigma) = \frac{M}{\kappa \pi} \left[ 
        \begin{pmatrix}
            1\\0\\0\\0
        \end{pmatrix}
        +
        \begin{pmatrix}
            0\\0\\0\\ \xi
        \end{pmatrix}
        \cos(\sigma) \right], \quad
        \rho^*_{\mu}(\sigma) = \frac{M}{\kappa \pi} 
        \begin{pmatrix}
            0\\1\\0\\0
        \end{pmatrix}
        \cos(\sigma). 
    \end{aligned}
\end{equation}
Let us perform the transition to the coordinate system oriented in the same way as the one, where the momenta of partons were initially defined (the ``$\sim$''-frame), using the matrix (\ref{eq:rotation_mat}):
\begin{equation}
    \label{eq:ic_cm_tilde_diff}
    \begin{aligned}
        \tilde{v}_{\mu}(\sigma) &= \frac{M}{\kappa \pi} \left[ 
        \begin{pmatrix}
            1\\
            R(\theta,\varphi)
            \begin{pmatrix}
            0\\0\\0
            \end{pmatrix}
        \end{pmatrix}
        +
        \begin{pmatrix}
            0\\
            R(\theta,\varphi)
            \begin{pmatrix}
            0\\0\\ \xi
            \end{pmatrix}
        \end{pmatrix}
        \cos(\sigma) \right]\\
        &=\frac{M}{\kappa \pi} \left[ 
        \begin{pmatrix}
            1\\0\\0\\0
        \end{pmatrix}
        + \xi
        \begin{pmatrix}
            0\\ \sin\varphi\sin\theta\\ -\cos\varphi\sin\theta\\ \cos\theta
        \end{pmatrix}
        \cos(\sigma) \right],
        \\
        \tilde{\rho}_{\mu}(\sigma) &= \frac{M}{\kappa \pi} 
        \begin{pmatrix}
            0\\ \cos\varphi\\ \sin\varphi\\ 0
        \end{pmatrix}
        \cos(\sigma). 
    \end{aligned}
\end{equation}
Let us introduce the notation:
\begin{equation}
    \label{eq:psilambda}
    \psi_{\mu} \equiv
    \begin{pmatrix}
        0\\ \sin\varphi\sin\theta\\ -\cos\varphi\sin\theta\\ \cos\theta
    \end{pmatrix}
    =
    \begin{pmatrix}
        0\\ \boldsymbol{\psi}
    \end{pmatrix}
    , \quad 
    \lambda_{\mu} \equiv
    \begin{pmatrix}
        0\\ \cos\varphi\\ \sin\varphi\\ 0
    \end{pmatrix}
    =
    \begin{pmatrix}
        0\\ \boldsymbol{\lambda}
    \end{pmatrix}
    .
\end{equation}
Then the equations (\ref{eq:ic_cm_tilde_diff}) take the form:
\begin{equation}
    \label{eq:ic_cm_tilde}
    \begin{aligned}
        \tilde{v}_{\mu}(\sigma) = (\kappa \pi)^{-1} M \left[ \delta_{0\mu} + \xi \psi_{\mu} \cos(\sigma) \right], \quad \tilde{\rho}_{\mu}(\sigma) = (\kappa \pi)^{-1} M \lambda_{\mu} \cos(\sigma). 
    \end{aligned}
\end{equation}
Next step is to use the Lorentz boost (\ref{eq:lorentz_boost_string}) to transform into the arbitrary reference frame where the string is moving with momentum $\boldsymbol{P}$. At first, one should note the simple relations: $-\boldsymbol{P}\boldsymbol{\psi} = P_{\mu}\psi^{\mu} \equiv (P\psi)$, $-\boldsymbol{P}\boldsymbol{\lambda} = (P\lambda)$. Substituting (\ref{eq:ic_cm_tilde}) into (\ref{eq:lorentz_boost_string}) gives the expression for the distributed momentum of the string:
\begin{equation*}
    \begin{aligned}
        p_0(\sigma) =  \frac{P_0 - \xi (P\psi) \cos(\sigma)}{\pi}, \quad \boldsymbol{p}(\sigma) = \frac{\boldsymbol{P}}{\pi} + \frac{\xi}{\pi} \left( M\boldsymbol{\psi} - \boldsymbol{P} \frac{(P\psi)}{P_0 + M} \right) \cos(\sigma).
    \end{aligned}
\end{equation*}
One can introduce the following notation
\begin{equation}
    \label{eq:chi}
    \chi_0 \equiv 1, \quad \boldsymbol{\chi} = \frac{\boldsymbol{P}}{P_0 + M}
\end{equation}
to define the vector of coefficients $\chi_{\mu}$, $\mu = 0,~\ldots,~3$, and to write the formula for the time and space components of the string initial velocity in a uniform way:
\begin{equation}
    \label{eq:foee1_ic_v}
    v_{\mu}(\sigma) = (\kappa \pi)^{-1} \left[ P_{\mu} + \xi \left( M\psi_{\mu} - (P\psi)\chi_{\mu} \right) \cos(\sigma) \right].
\end{equation}
The same could be done for the function $\tilde{\rho}_{\mu}(\sigma)$ from (\ref{eq:ic_cm_tilde}) to obtain
\begin{equation}
    \label{eq:foee1_ic_rho}
    \rho_{\mu}(\sigma) = (\kappa \pi)^{-1} \left( M\lambda_{\mu} - (P\lambda)\chi_{\mu} \right) \cos(\sigma).
\end{equation}

The formulae (\ref{eq:foee1_ic_v}), (\ref{eq:foee1_ic_rho}) define the initial conditions for the FOEE(1)-string in the arbitrary frame of reference, where the numeric coefficients are calculated using Eqs. (\ref{eq:psilambda}), (\ref{eq:chi}) and (\ref{eq:thetaphi}). This method is applicable for any string with non-zero mass $M$ and for any arbitrary frame of reference. Although we have restricted ourselves to the case of a string with only one eigenharmonic, it is nonetheless a universal way to deal with the massive quark-gluon string in $3+1$ dimensions. 

An important thing to note is that only the 4-momenta of two partons between which the string is stretched are required to completely define the string motion. Their relative positions in coordinate space are not important for this method and can only influence the sign of the string rotation signature $\xi$, which means that it could be randomly selected as the relative position of two partons in the transverse plane of colliding particles is invariant of the orientation of the coordinate system.

Now one can derive the formula for the Fourier amplitudes. Let us substitute Eqs. (\ref{eq:foee1_ic_v}), (\ref{eq:foee1_ic_rho}) into the standard expressions:
\begin{equation}
    \label{eq:foee1_famp}
    \begin{aligned}
        \alpha_{\pm 1\mu} &= \sqrt{\frac{\kappa}{\pi}} \int_{0}^{\pi} d\sigma \left[ v_{\mu}(\sigma) \mp i\rho_{\mu}(\sigma) \right] \cos(\sigma)\\
        &= \frac{1}{2\sqrt{\kappa \pi}} \left[ \xi \left(M\psi_{\mu} - (P\psi)\chi_{\mu} \right) \mp i \left( M\lambda_{\mu} - (P\lambda)\chi_{\mu} \right) \right]\\
        &= \frac{1}{2\sqrt{\kappa \pi}} \left[M(\xi \psi_{\mu} \mp i \lambda_{\mu}) - P_{\nu}(\xi \psi^{\nu} \mp i \lambda^{\nu})\chi_{\mu}\right]\\
        &= \frac{M\omega_{\mu}^{\pm} - (P\omega^{\pm})\chi_{\mu}}{2\sqrt{\kappa \pi}},
    \end{aligned}
\end{equation}
where vectors $\omega_{\mu}^{\pm}$ are introduced:
\begin{equation}
    \label{eq:omega}
    \omega_{\mu}^{\pm} \equiv
    \begin{pmatrix}
        0\\ \xi\sin\varphi \sin\theta \mp i \cos\varphi \\ -\xi \cos\varphi \sin\theta \mp i \sin\varphi \\ \xi \cos\theta
    \end{pmatrix}
    .
\end{equation}
One can check that Fourier amplitudes (\ref{eq:foee1_famp}) satisfy the Virasoro conditions (\ref{eq:virasoro_cond}):
\begin{equation*}
    (\alpha_{\pm1\mu})^2=\alpha_{\pm1}\alpha_0=2\alpha_{-1}\alpha_1 + (\alpha_{0\mu})^2=0,
\end{equation*}
where $\alpha_{0\mu} \equiv {P_{\mu}}/{\sqrt{\kappa \pi}}$ as usual. The fulfillment of the Virasoro conditions for the Fourier amplitudes (\ref{eq:foee1_famp}) proves that the Lorentz boost for the string (\ref{eq:lorentz_boost_string}) is a correctly defined transformation.

It is time to finally obtain the expression for the coordinates of the massive string moving in $3+1$ dimensions. We have:
\begin{equation*}
    x_{\mu}(\tau,\sigma) = Q_{\mu} + \frac{P_{\mu}\tau}{\kappa \pi} + \frac{i}{2\kappa \pi} \left( -\left[ M\omega_{\mu}^- - (P\omega^-)\chi_{\mu} \right] e^{i\tau} + \left[ M\omega_{\mu}^+ - (P\omega^+)\chi_{\mu} \right] e^{-i\tau} \right) \cos(\sigma).
\end{equation*}
Let us define
\begin{equation}
    \label{eq:Omega}
    \omega_{\mu}^+e^{-i\tau} - \omega_{\mu}^-e^{i\tau} = -2i
    \left[ \lambda_\mu \cos(\tau) + \xi \psi_\mu \sin(\tau) \right]
    \equiv -2i\Omega_{\mu}(\tau)
\end{equation}
and calculate
\begin{equation*}
    Q_{\mu} = \int_{0}^{\pi} d\sigma \rho_{\mu}(\sigma) =  \frac{M\lambda_{\mu} - (P\lambda)\chi_{\mu}}{\kappa \pi}  \int_{0}^{\pi} \cos(\sigma)d\sigma = 0.
\end{equation*}
Then we obtain the final formula for the coordinates of the FOEE(1)-string:
\begin{equation}
    \label{eq:foee1_x}
    x_{\mu}(\tau,\sigma) = (\kappa \pi)^{-1} \left[ P_{\mu}\tau + \left[ M\Omega_{\mu}(\tau) - (P\Omega(\tau))\chi_{\mu} \right] \cos(\sigma) \right].
\end{equation}
The formula for the velocity of the string is the following:
\begin{equation}
    \label{eq:foee1_xdot}
    \dot{x}_{\mu}(\tau, \sigma) = (\kappa \pi)^{-1} \left[ P_{\mu} + \xi \left[ M\Lambda_{\mu}(\tau) - (P\Lambda(\tau))\chi_{\mu} \right] \cos(\sigma) \right],
\end{equation}
where
\begin{equation}
    \label{eq:Lambda}
    \Lambda_\mu(\tau) \equiv \xi \dot{\Omega}_\mu(\tau) =
    \psi_\mu \cos(\tau) - \xi \lambda_\mu \sin(\tau)
\end{equation}
The derivative $\partial x /\partial\sigma$ is
\begin{equation}
    \label{eq:foee1_xprime}
    x^{\prime}_{\mu}(\tau,\sigma) = -(\kappa \pi)^{-1} \left[ M\Omega_{\mu}(\tau) - (P\Omega(\tau))\chi_{\mu} \right] \sin(\sigma).
\end{equation}

One can check that the conditions on the tangent vectors to the world sheet of the string are fulfilled for the expressions (\ref{eq:foee1_xdot}), (\ref{eq:foee1_xprime}):
\begin{equation*}
    (\dot{x}_{\mu}(\tau,\sigma))^2 = \frac{M^2}{(\kappa \pi)^2} \sin^2(\sigma) \ge 0, \quad (x^{\prime}_{\mu}(\tau, \sigma))^2 = -\frac{M^2}{(\kappa \pi)^2} \sin^2(\sigma) \le 0.
\end{equation*}
The expressions also match those obtained in the center-of-mass system, which once again proves the Lorentz invariance of the proposed approach.

\begin{figure}[htbp]
    \centering
    \includegraphics[width=.6\textwidth]{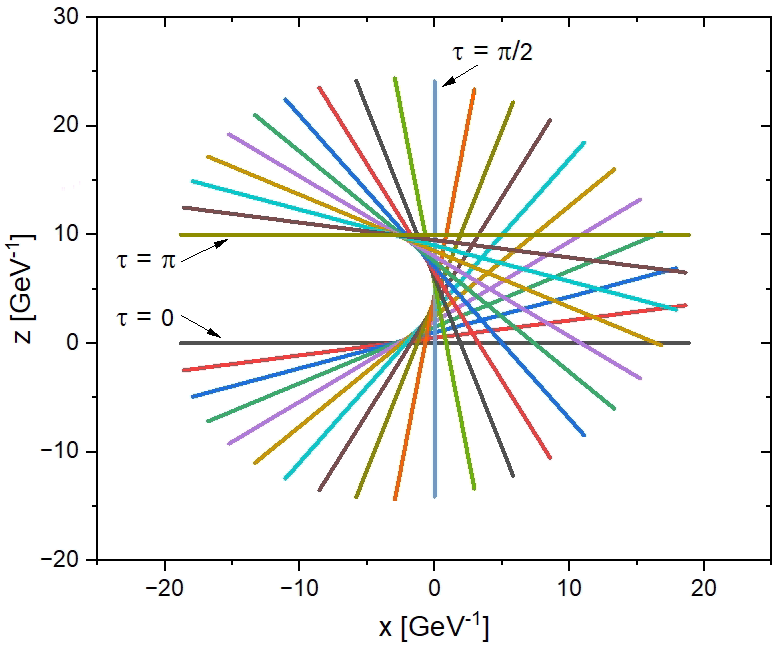}
    \caption{Motion of the FOEE(1)-string defined with total momentum $P_z = 2$ GeV.\label{fig:fig2}}
\end{figure}

Let us give an example of a string defined using the initial conditions (\ref{eq:foee1_ic_v}), (\ref{eq:foee1_ic_rho}). Let the string model a system of two partons, the momenta of which are $P_{1\mu} = (6,~3,~0,~1)$ GeV and $P_{2\mu} = (6,-3,~0,~1)$ GeV, respectively. The mass of the string is $M\approx12$ GeV. The result of calculating the motion according to (\ref{eq:foee1_x}) with a step $\Delta \tau = \pi / 20$ is shown in figure \ref{fig:fig2}. It is evident that the string still rotates as a rigid rod, but now the motion of the system as a whole is also added. The displacement along the $Z$ axis is $10 \text{ GeV}^{-1}$, as expected, since the center of mass of the string in time $\tau=\pi$ will pass $P_z \pi / (\kappa \pi) = 2 / 0.2 = 10 \text{ GeV}^{-1}$ (the generally accepted value of the string tension coefficient $\kappa \approx 0.2 \text{ GeV}^2$ is used).

\begin{figure}[htbp]
    \centering
    \includegraphics[width=.6\textwidth]{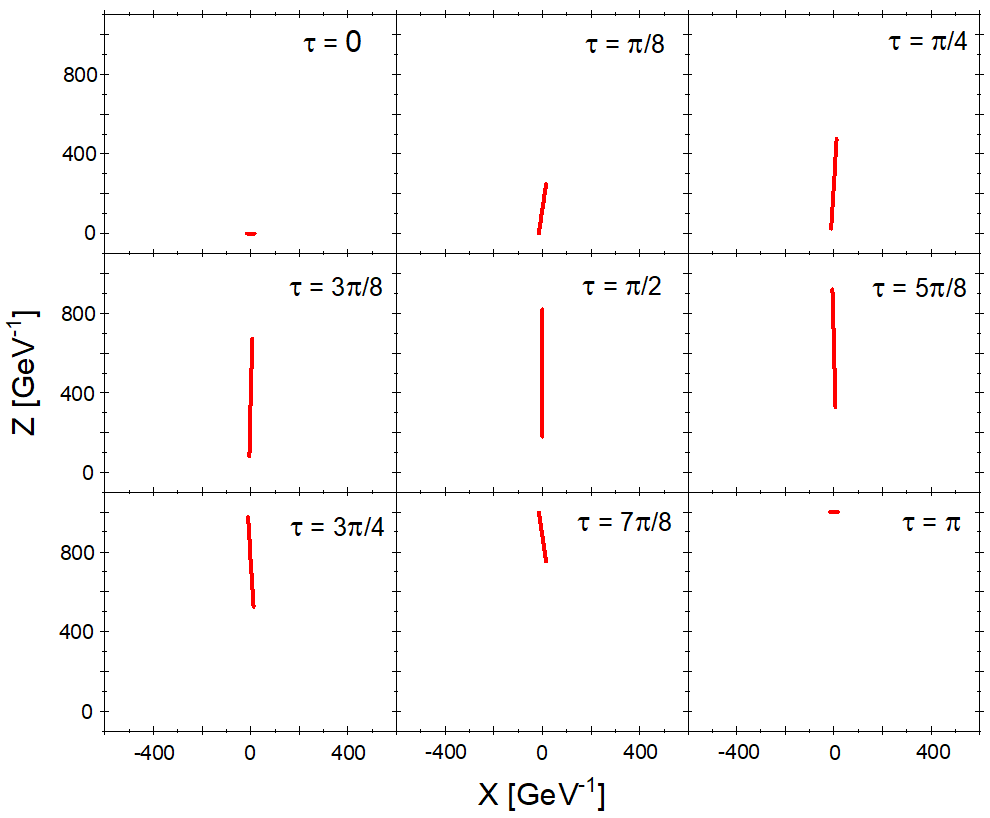}
    \caption{Frame-by-frame motion of the FOEE(1)-string defined with total momentum $P_z = 200$ GeV.\label{fig:fig4}}
\end{figure}

It is useful to provide an example of a string moving with huge momentum. Figure \ref{fig:fig4} demonstrates one half-cycle ($\tau \in [0,\pi]$) of the string defined with $P_z=200$ GeV and $M=10$ GeV. One can clearly see that, due to one of the ends of the string moving against the direction of the total string velocity  at $\tau \sim \pi/2$, the string stretches strongly along the $Z$ direction.

\subsection{Generalization to the string with oscillation defined by eigenharmonic of arbitrary order}
\label{sec:nustr}
By definition, the FOEE(1)-string can have only one order in the expansion in eigenfunctions with a non-zero amplitude value (in addition to the zeroth order). However, this does not mean that functions of the form (\ref{eq:foee1_ic_base}) are the only possible ones. Indeed, due to the arbitrariness of the choice of the $\sigma$ parameter values, instead of the first-order eigenfunction $\cos(\sigma)$, one can use any higher oscillation $\cos(\nu \sigma)$ , where $\nu$ is an arbitrary natural number, $\nu > 1$. If the initial data are taken in the form:
\begin{equation}
    \label{eq:nustr_ic_cm_base}
    v_{\mu}(\sigma) = a_{\mu} + b_{\mu}\cos(\nu \sigma), \quad \rho_{\mu}(\sigma) = c_{\mu} + d_{\mu} \cos(\nu \sigma),
\end{equation}
then, among all the Fourier amplitudes $\alpha_{n\mu}$, only the amplitudes $\alpha_{\pm \nu \mu}$ will have a non-zero value. This means that the general form of the FOEE system will not change (only the coefficients in the equations might differ).

Repeating for the initial conditions (\ref{eq:nustr_ic_cm_base}) the actions similar to those that led to the system (\ref{eq:foee1_virasoro}) and Eq. (\ref{eq:foee1_consM}) for the initial data (\ref{eq:foee1_ic_base}), we obtain the FOEE(1) system for a string with eigenoscillation of order $\nu$ ($\nu \ge 1$):
\begin{equation}
    \label{eq:nustr_foeesyst}
    \left\{
    \begin{aligned}
        b^2 - \nu^2 d^2 &= 0\\
        bd = bP = dP &= 0\\
        b^2 + d^2 + \frac{2P^2}{(\kappa \pi)^2} &= 0\\
        a_{\mu} &= \frac{P_{\mu}}{\kappa \pi}\\
        \mathcal{M}_{\mu \nu} = c_{\mu}P_{\nu} - c_{\nu}P_{\mu} &+ \frac{\kappa \pi}{2} \left( d_{\mu}b_{\nu} - d_{\nu}b_{\mu} \right).
    \end{aligned}
    \right.
\end{equation}
As one can see, only the first equation has changed. It is now trivial to solve this system in the center-of-mass frame. It also yields a relation between the string mass and spin similar to (\ref{eq:spin_mass1}):
\begin{equation}
    \label{eq:nustr_spinmass}
    2 \kappa \pi \nu J = M^2.
\end{equation}
Note that condition (\ref{eq:nustr_spinmass}), unlike (\ref{eq:spin_mass1}), does not fix the spin value of the string with defined mass, since the order of the eigenfunction $\nu$, as previously stated, is an arbitrary natural number. Thus, the FOEE(1)-string with a given mass has a discrete spectrum of possible spin values that depend on the order of the eigenoscillation with a non-zero amplitude. Thanks to this, in practice it is possible to select the value of $\nu$ in the case when the string spin is assumed to be known based on some other assumptions. For example, if an FOEE(1)-string with a mass of $100$ GeV ``must'' have a spin of $J=1/2$, the order of its non-zero oscillation should be taken equal to
\begin{equation*}
    \nu = \frac{M^2}{2\kappa\pi J} \approx 0.8\frac{10^4}{0.5}=16~000.
\end{equation*}
Of course, using the formula (\ref{eq:nustr_spinmass}) it is impossible to choose such a value of $\nu$ to perfectly match the integer or half-integer quantum spin numbers. Yet, it allows the existence of massive strings with not very high classical angular momentum within the framework of the considered theory. As the relative position of the ends of the string is not known in this approach, the sign of the projection of angular momentum in the rotation plane is arbitrary; let it be equal to $\xi$.

Thus, we can require the massive quark-gluon string to have: large values of eigenharmonic order $\nu$ for the case of $e^+e^-$ collisions where one can expect the angular momentum of the string to be defined by the spins of the end-point partons only; moderate $\nu$ values for hadron-hadron interactions where the impact parameter of colliding partons can add significant rotation; small $\nu$ values for ion collisions. Thus, a natural way to overcome the universality of the quark-gluon string fragmentation is achieved.

The case of a string with zero spin requires $\nu \rightarrow\infty$; the behavior of the functions that define the coordinates and velocity of the string will be considered separately.

Let us note an important feature of the string defined with $\nu > 1$. There is now a set of points $\sigma^{\text{joint}}_n = n \pi /\nu$ ($n$ is an integer, $0<n<\nu$) at which the oscillation function $\cos(\nu\sigma)$ reaches its extreme values corresponding to the string end-points in $x_{\mu}$ coordinate space. This means that a string (a rigid rod, as we have seen) of this configuration can be interpreted as $\nu -1$ times ``folded'' FOEE(1)-string with $\nu=1$, and the points $\sigma^{\text{joint}}_n$ ($n=1,~\ldots,~\nu-1$) are the ``joints'' of this ``folded'' rod. Also note that if $\nu$ is even, the point $\sigma = \pi$ corresponds to the same coordinates in space as $\sigma = 0$. Thus, one should consider that the other end point of the string is placed at $\sigma = \pi/2$. By ``folding'' the string we can achieve the reduction of its angular momentum (as we decrease its length), while the mass of the string remains the same. An ambiguity in interpreting how the string is attached to partons appears, but it is irrelevant to discuss it within the framework of the considered model, as the realistic configuration of the QCD field between two partons is beyond the initial postulates of the relativistic string model.

Hereinafter, the considered class of massive relativistic strings, defined by the FOEE method with a single non-zero Fourier amplitude of the eigenoscillation of order $\nu \ge 1$, rotation signature $\xi$, and modeling the simplest parton-antiparton system (one-segment string), will be denoted as $1_\nu^\xi$-strings. It should be noted that the generalization to the case of strings consisting of a larger number of partons ($N$-segment strings) is non-trivial within the framework of considered approach and is worthy of a separate article on its own.

The initial conditions for the $1_\nu^\xi$-string defined in the center-of-mass are:
\begin{equation}
    \label{eq:nustr_ic_cm}
    v_{\mu}(\sigma) = (\kappa \pi)^{-1} M \left[ \delta_{0\mu} + \xi \delta_{3\mu} \cos(\nu\sigma) \right], \quad \rho_{\mu}(\sigma) = (\nu\kappa \pi)^{-1}M \delta_{1_\mu} \cos(\nu \sigma).
\end{equation}
Calculating Fourier amplitudes for (\ref{eq:nustr_ic_cm}) gives
\begin{equation}
    \label{eq:nustr_cm_famp}
    \alpha_{\pm \nu \mu} = \sqrt{\frac{\kappa}{\pi}} \frac{M}{\kappa \pi} \left[ \xi \delta_{3\mu} \mp i\nu \frac{\delta_{1\mu}}{\nu} \right] \int_{0}^{\pi}\cos^2(\nu \sigma) d\sigma = \frac{M\left( \xi \delta_{3\mu} \mp i \delta_{1\mu} \right)}{2\sqrt{\kappa \pi}},
\end{equation}
which matches the case of $\nu=1$. So, the fulfillment of the Virasoro conditions is guaranteed. The formula for the coordinates of the string takes the form:
\begin{equation}
    \label{eq:nustr_cm_x}
    x_{\mu}(\tau,\sigma) = (\kappa \pi)^{-1} M \left[ \delta_{0\mu}\tau + \nu^{-1} \left[ \xi \delta_{3\mu}\sin(\nu\tau) + \delta_{1\mu}\cos(\nu\tau) \right] \cos(\nu\sigma) \right].
\end{equation}
From a comparison of formula (\ref{eq:nustr_cm_x}) with (\ref{eq:foee1_cm_solution}), it is obvious that the conditions on the tangent vectors to the world sheet of the string (\ref{eq:tang_vec}) are satisfied.

Let the orientation of the string rotation plane be determined again by the angles $\theta$, $\varphi$ through the matrix (\ref{eq:rotation_mat}). Then the initial conditions for the string defined in this arbitrary system are described by the following expressions:
\begin{equation}
    \label{eq:nustr_ic}
    \begin{aligned}
        v_{\mu}(\sigma) &= (\kappa \pi)^{-1} \left[ P_{\mu} + \xi\left( M\psi_{\mu} - (P\psi)\chi_{\mu} \right) \cos(\nu \sigma) \right],\\
        \rho_{\mu}(\sigma) &= (\nu\kappa \pi)^{-1} \left( M\lambda_{\mu} - (P\lambda)\chi_{\mu} \right) \cos(\nu \sigma),
    \end{aligned}
\end{equation}
where coefficients $\psi_{\mu}$, $\lambda_{\mu}$ and $\chi_{\mu}$ are defined in Eqs. (\ref{eq:psilambda}) and (\ref{eq:chi}) as before. The Fourier amplitudes $\alpha_{\pm\nu\mu}$ in the case of a string defined by the initial conditions (\ref{eq:nustr_ic}) are given by the expression:
\begin{equation}
    \label{eq:nustr_famp}
    \alpha_{\pm\nu\mu} = \frac{M\omega_{\mu}^{\pm} - (P\omega^{\pm})\chi_{\mu}}{2\sqrt{\kappa \pi}},
\end{equation}
where
\begin{equation}
    \label{eq:nustr_omega}
    \omega_{\mu}^{\pm} \equiv
    \begin{pmatrix}
        0\\ \xi\sin\varphi \sin\theta \mp i \cos\varphi \\ -\xi \cos\varphi \sin\theta \mp i \sin\varphi \\ \xi \cos\theta
    \end{pmatrix}
\end{equation}
One can see that there is no difference between (\ref{eq:foee1_famp}) and (\ref{eq:nustr_famp}). So, all conditions on the Fourier amplitudes imposed by the Virasoro conditions are certainly satisfied. The solution to the Cauchy boundary-value problem for the motion of the relativistic string is given by a formula:
\begin{equation}
    \label{eq:nustr_x}
    x_{\mu}(\tau, \sigma) = (\kappa \pi)^{-1} \left[ P_{\mu}\tau + \nu^{-1} \left[ M\Omega_{\mu}(\tau, \nu) - (P\Omega(\tau,\nu))\chi_{\mu} \right] \cos(\nu\sigma) \right],
\end{equation}
where
\begin{equation}
    \label{eq:nustr_Omega}
    \Omega_{\mu}(\tau, \nu) \equiv -\frac{i}{2}\left( \omega_{\mu}^+e^{-i\nu\tau} - \omega_{\mu}^-e^{i\nu\tau} \right) = \lambda_\mu \cos(\nu\tau) + \xi \sin(\nu\tau)
    .
\end{equation}
The velocity of the string is described by the expression:
\begin{equation}
    \label{eq:nustr_xdot}
    \dot{x}_{\mu}(\tau, \sigma) = (\kappa \pi)^{-1} \left[ P_{\mu} + \xi \left[ M\Lambda_{\mu}(\tau, \nu) - (P\Lambda(\tau,\nu))\chi_{\mu} \right] \cos(\nu\sigma) \right],
\end{equation}
with
\begin{equation}
    \label{eq:nustr_Lambda}
    \Lambda_{\mu}(\tau,\nu) \equiv \xi \nu^{-1} \dot{\Omega}_{\mu}(\nu,\tau) =  -\frac{\xi}{2} \left( \omega_{\mu}^+e^{-i\nu\tau} + \omega_{\mu}^-e^{i\nu\tau} \right).
\end{equation}
The expression for the derivative $x_{\mu}^{\prime}(\tau,\sigma)$ is:
\begin{equation}
    \label{eq:nustr_xprime}
    x_{\mu}^{\prime}(\tau,\sigma) = -(\kappa \pi)^{-1} \left[ M\Omega_{\mu}(\tau, \nu) - (P\Omega(\tau,\nu))\chi_{\mu} \right] \sin(\nu\sigma).
\end{equation}
By comparing (\ref{eq:nustr_x}) with (\ref{eq:foee1_x}), it is easy to verify that (\ref{eq:nustr_xdot}) and (\ref{eq:nustr_xprime}) meet the conditions on the tangent vectors to the world sheet of the string.

From expression (\ref{eq:nustr_x}) it is evident that, when $\nu \rightarrow \infty$, the string shrinks to a relativistic point-like particle moving with a 4-momentum $P_\mu$ and does not possess angular momentum according to (\ref{eq:nustr_spinmass}). The expression (\ref{eq:nustr_xdot}) in this case is only formally needed to maintain consistency in the description of the characteristics of the string in terms of the quantities distributed along its length. Note that in this case, the string, which was a point at the initial moment in time, remains a point at any other moment in time. This shows a clear divergence from the usually accepted Lund or Morris approaches, where the string evolves from a point into a stretched object.

\subsection{The algorithm to calculate the motion of the FOEE(1) relativistic string}
\label{sec:alg}
Let us summarize the obtained results in the form of an algorithm for calculating the motion of a $1_\nu^\xi$-string in an arbitrary frame of reference, using which a coupled system of two partons is modeled.

Let the 4-momenta of two partons (e.g., a quark and an antiquark) combined into a string be given. Then one should perform the following sequence of actions.
\begin{enumerate}
    \item Define the invariant mass of the parton system (the mass of the string) $M$ as the square root of the total 4-momentum of the system $P_{\mu}$. Perform the transition to the center-of-mass system of two partons using the inverse Lorentz transformations:
    \begin{equation}
        \label{eq:lorentz_reverse}
        \omega_i^* = \frac{P_0\omega_i - \boldsymbol{P}\boldsymbol{p}_i}{M}, \quad \boldsymbol{p}^*_i = \boldsymbol{p}_i - \boldsymbol{P} \frac{\omega_i + \omega_i^*}{P_0 + M},
    \end{equation}
    where $\omega_i$, $\boldsymbol{p}_i$ are the energy and momentum of the $i$-th particle in the original system, $i=1,~2$.
    
    \item Knowing the coordinates of the 4-momenta of partons in the center-of-mass system, calculate the elements of the rotation matrix (\ref{eq:rotation_mat}) using (\ref{eq:thetaphi}). 
    
    \item Set the signature of the string rotation (as it was explained, the sign of (\ref{eq:xi}) is, in fact, an arbitrary value). Choose the order $\nu$ of the eigenoscillation of the string.
    
    \item Calculate the coordinates of the vectors $\psi_{\mu}$, $\lambda_{\mu}$ and $\chi_{\mu}$ using (\ref{eq:psilambda}) and (\ref{eq:chi}), respectively. Then the functions of the initial conditions of the Cauchy boundary-value problem $\rho_\mu(\sigma) \equiv x_\mu(0,\sigma)$, $v_\mu (\sigma) \equiv \dot{x}_\mu(0,\sigma)$ are given according to (\ref{eq:nustr_ic}).
    
    \item Calculate the Fourier amplitudes of the order $\pm\nu$ of expansion using the expressions (\ref{eq:nustr_famp}), (\ref{eq:nustr_omega}).
    
    \item To calculate the coordinates of the string $x_\mu (\tau, \sigma)$ at an arbitrary moment of time, one should first obtain the coordinates of the vector $\Omega_\mu (\tau,\nu)$ according to (\ref{eq:nustr_Omega}), and then apply the formula (\ref{eq:nustr_x}). To calculate the velocity of the string points $\dot{x}_\mu(\tau, \sigma)$, calculate $\Lambda_\mu(\tau,\nu)$ using (\ref{eq:nustr_Lambda}) and use the expression (\ref{eq:nustr_xdot}). Note that the distributed momentum of the string is simply $p_\mu(\tau,\sigma) = \kappa \dot{x}_\mu(\tau,\sigma)$.
\end{enumerate}

Thus, a complete and consistent with the Virasoro conditions method for calculating the motion of a massive quark-gluon string with angular momentum is constructed.

\section{String fragmentation}
\label{sec:fragm}
Since within the framework of considered model hadronization is represented as an iterative process of string breaking apart, it is necessary to derive the basics of the fragmentation of $1_\nu^\xi$-strings. I should note that only general aspects will be discussed here, as a complete space-time picture of string fragmentation is strongly influenced by the adopted ideology of the transition from string state to hadron state. Here, the focus will be on the mathematical apparatus that allows one to correctly describe the breaking of the string and the properties of newly created daughter strings.

Let us begin with an expression for the invariant area of the string world sheet. When using the orthonormal gauge (\ref{eq:ONG}) the expression for the invariant area $A$, which has been covered by the string by the time $\tau$, can be calculated as follows:
\begin{equation}
    \label{eq:invA_ong_gen}
    A(\tau) \equiv \int_{0}^{\tau}d\tau^{\prime} \int_{0}^{\pi} d\sigma \sqrt{(\dot{x}x^{\prime})^2-\dot{x}^2x^{\prime2}} = \int_{0}^{\tau}d\tau^{\prime} \int_{0}^{\pi} \dot{x}^2(\tau^\prime, \sigma) d\sigma.
\end{equation}
We can use either the expression (\ref{eq:nustr_cm_x}) or (\ref{eq:nustr_xdot}) to obtain the formula for the case of $1_\nu^\xi$-string:
\begin{equation}
    \label{eq:invA_ong}
    A(\tau) = \int_{0}^{\tau} d\tau^\prime \int_{0}^{\pi} \frac{M^2}{(\kappa \pi)^2} \sin^2(\nu \sigma) d\sigma = \frac{M^2 \tau}{2 \pi \kappa^2}.
\end{equation}
According to Eq. (\ref{eq:AreaDec}), the probability that string did not decay by the moment in time $\tau$ is described by:
\begin{equation}
    \label{eq:prob_to_not_decay}
    P(\tau) \propto \exp{\left(-P_0\frac{M^2\tau}{2\pi \kappa} \right)},
\end{equation}
where $P_0$ is a dimensionless decay constant (in the absence of any theoretical insights into the string decay cross-sections, it is tuned as a free parameter of the model). If string fragmentation is governed by the area decay law (\ref{eq:AreaDec}), the moment in time $\tau$ when the string breaks should be sampled according to (\ref{eq:prob_to_not_decay}). As the $\sigma$ coordinate of the break point is not explicitly present in the decay law (\ref{eq:prob_to_not_decay}), at this point it should be considered a uniformly distributed random quantity $\sigma \in [0, \pi]$.

Let us consider an arbitrary break point $(\tau_{\text{break}}, \sigma_{\text{break}}) \equiv (\tau^*, \sigma^*)$. The total momenta of the string fragments (daughter strings) are calculated as:
\begin{equation}
    \label{eq:frag_mom_gen}
    \begin{aligned}
        P_{1\mu} &= \kappa \int_{0}^{\sigma^*} d\sigma \dot{x}_{\mu}(\tau^*,\sigma) = \frac{P_{\mu}\sigma^*}{\pi} + \left[ M\Lambda_{\mu}(\tau^*,\nu) - (P\Lambda(\tau^*,\nu))\chi_{\mu} \right] \frac{\sin(\sigma^*)}{\nu\pi},\\
        P_{2\mu} &= \kappa \int_{\sigma^*}^{\pi} d\sigma \dot{x}_{\mu}(\tau^*,\sigma) = \frac{P_{\mu}(\pi-\sigma^*)}{\pi} - \left[ M\Lambda_{\mu}(\tau^*,\nu) - (P\Lambda(\tau^*,\nu))\chi_{\mu} \right] \frac{\sin(\sigma^*)}{\nu\pi}.
    \end{aligned}
\end{equation}
Now using (\ref{eq:frag_mom_gen}) one can obtain their masses:
\begin{equation}
    \label{eq:frag_mass_gen}
    M_1^2 = P_1^2 = \frac{M^2}{\pi^2} \left( \sigma^{*2} - \frac{\sin^2(\nu\sigma^*)}{\nu^{2}} \right), \quad M_2^2 = \frac{M^2}{\pi^{2}} \left( (\pi - \sigma^{*})^2 - \frac{\sin^2(\nu\sigma^*)}{\nu^{2}} \right).
\end{equation}
Note that in Eq. (\ref{eq:frag_mass_gen}) the quantities are strictly positive: $M_1^2 > 0$, $\sigma^* > 0$,  and $M_2^2 > 0$, $\sigma^* < \pi$.

\subsection{The boundary-value problem for the motion of a daughter string}
\label{sec:daughter_prob}

The formulae (\ref{eq:frag_mom_gen}) and (\ref{eq:frag_mass_gen}) have a general meaning. They allow calculation of the momentum and mass of the string pieces and do not depend on actual breaking of the string. To proceed with developing the framework for string fragmentation, we will consider all necessary restrictions arising from the strict theory. In the classical Nambu-Goto string theory, such a process as string breaking is not described (that would require the addition of the corresponding term to the action (\ref{eq:NGaction})). Therefore, we will define a new special scheme for this. It is clear that the coordinates of the string fragments at the breaking time must, as a whole, repeat the shape of the mother string. The same applies to the velocity function $\dot{x}_{\mu}(\tau, \sigma)$. We will assume that immediately at the time of breaking, both daughter strings are no longer connected by the chromoelectric field. So, for each of them, the boundary conditions of a string with free ends can be applied.

Since at the time of break all parameters of the string fragments are defined by functions of the variable $\sigma$ only, to formulate the problem of the motion of the daughter strings, it is convenient to assume that the parameter $\tau$ is again measured starting from zero ($\tau > 0$, and the value $\tau = 0$ corresponds to the time of break of the mother string). The interval of values of the parameter $\sigma$ will now be defined differently: $\sigma \in [\sigma_1, \sigma_2]$, where $\sigma_{1,2}$ are some boundary values, $0 \le \sigma_1 < \sigma_2 \le \pi$.

To avoid confusion between mother-and-daughter strings and between string fragments of the same generation, we will add two indices to all calculated values. Let the index $g$ denote the generation to which a given string belongs. For primary strings $g = 0$. The index $j$ will denote the ordinal number of the string in generation. We will number the strings from $j = 1$ to $j=2^g$ from the end $\sigma = \sigma_{gj1}$ to the end $\sigma = \sigma_{gj2}$.

Then the statement of the problem for the motion of a string fragment with number $j$ in generation $g$ is the following:
\begin{equation}
    \label{eq:fragm_CauchyProb}
    \begin{aligned}
         \ddot{x}_{gj\mu}-x^{\prime\prime}_{gj\mu} &= 0,\quad \sigma \in\left[\sigma_{gj1}, \sigma_{gj2}\right], \quad \tau>0, \quad \mu=0,~\ldots,~3;\\
         x^{\prime}_{gj\mu}(\tau,\sigma_{gj1}) & =x^{\prime}_{gj\mu}(\tau,\sigma_{gj2}) = 0;\\
         x_{gj\mu}(0,\sigma) & = \rho_{gj\mu}(\sigma) \equiv x_{(g-1)k\mu}(\tau_{g-1}^*,\sigma),\\
         \dot{x}_{gj\mu}(0,\sigma) &= v_{gj\mu}(\sigma) \equiv \dot{x}_{(g-1)k\mu}(\tau_{g-1}^*,\sigma),
    \end{aligned}
\end{equation}
where the number $k$ denotes the number of the mother string in generation $g-1$.

We will, as before, look for the solution in the form:
\begin{equation*}
    x_{gj\mu}(\tau, \sigma) = T_{gj\mu}(\tau)u_{gj}(\sigma).
\end{equation*}
Let us write the equalities obtained after the separation of variables in the following way:
\begin{equation*}
    \frac{\ddot{T}_{gj\mu}(\tau)}{T_{gj\mu}(\tau)} = \frac{u_{gj}^{\prime \prime}(\sigma)}{u_{gj}(\sigma)} = - \left( \frac{\sigma_{gj2} - \sigma_{gj1}}{\pi} \right)^2 \omega_{gj}^2, \quad \tau > 0, \quad \sigma \in [\sigma_{gj1}, \sigma_{gj2}].
\end{equation*}
The choice of the numeric coefficient before $\omega_{gj}^2$ is made for convenience. The statement of the Sturm-Liouville problem for the eigenvalues $\omega_{gjn}$ and for the corresponding eigenfunctions $u_{gjn}(\sigma)$ is:
\begin{equation}
    \label{eq:fragm_sturm}
    \begin{aligned}
        u_{gjn}^{\prime \prime} (\sigma) + \left( \frac{\sigma_{gj2} - \sigma_{gj1}}{\pi} \right)^2 \omega_{gjn}^2 u_{gjn}(\sigma) &= 0, \quad \sigma \in [\sigma_{gj1}, \sigma_{gj2}], \\
        u_{gjn}^{\prime}(\sigma_{gj1}) = u_{gjn}^{\prime}(\sigma_{gj2}) &= 0.
    \end{aligned}
\end{equation}
We will write the general solution to the problem (\ref{eq:fragm_sturm}) as:
\begin{equation}
    \label{eq:fragm_u_gen}
    u_{gjn}(\sigma) = A_{gjn}\cos{\left( \omega_{gjn}\frac{\pi (\sigma - \sigma_{gj1})}{\sigma_{gj2} - \sigma_{gj1}} \right)} + B_{gjn}\sin{\left( \omega_{gjn}\frac{\pi (\sigma - \sigma_{gj1})}{\sigma_{gj2} - \sigma_{gj1}} \right)}, \quad n = 1,~2,~\ldots ~.
\end{equation}
It is easy to verify that the functions (\ref{eq:fragm_u_gen}) actually satisfy the differential equation (\ref{eq:fragm_sturm}). The essence of choosing the type of function of the argument $\sigma$ inside the brackets in Eq. (\ref{eq:fragm_u_gen}) is that the functions (\ref{eq:fragm_u_gen}) behave at their boundaries $\sigma = \sigma_{gj1,2}$ in the same way as functions of the form (\ref{eq:Aufunc_gen}) at their boundaries $\sigma = 0$, $\sigma =\pi$. Therefore, taking into account the boundary conditions of problem (\ref{eq:fragm_sturm}), we immediately obtain:
\begin{equation}
    \label{eq:fragm_w_u}
    \begin{aligned}
        \omega_{gjn} = n, \quad u_{gjn}(\sigma) &= \cos{\left( \pi n \frac{\sigma - \sigma_{gj1}}{\sigma_{gj2} - \sigma_{gj1}} \right)}, \quad n = 1,~2,~\ldots,\\
        u_{gj0}(\sigma) &\equiv 1.
    \end{aligned}
\end{equation}
One can effortlessly derive the orthogonality property of eigenfunctions (\ref{eq:fragm_w_u}) using obvious variable substitution:
\begin{equation}
    \label{fragm_ort}
    \begin{aligned}
        \int_{\sigma_{gj1}}^{\sigma_{gj2}} u_{gjn}(\sigma) u_{gjm}(\sigma) d\sigma & = \int_{\sigma_{gj1}}^{\sigma_{gj2}} \cos{\left( \pi n \frac{\sigma - \sigma_{gj1}}{\sigma_{gj2} - \sigma_{gj1}} \right)} \cos{\left( \pi m \frac{\sigma - \sigma_{gj1}}{\sigma_{gj2} - \sigma_{gj1}} \right)} d\sigma\\
        &= \frac{\sigma_{gj2} - \sigma_{gj1}}{\pi} \int_{0}^{\pi} \cos(nx)\cos(mx)dx\\
        &= \frac{\sigma_{gj2} - \sigma_{gj1}}{\pi} \frac{\pi}{2} \delta_{nm}.
    \end{aligned}
\end{equation}

We will write the general solution for function $T_{gjn\mu}(\tau)$ as:
\begin{equation}
    \label{eq:fragm_T_gen}
    \begin{aligned}
        T_{gjn\mu}(\tau) & = C_{gjn\mu} \cos{\left( \frac{n \pi \tau}{\sigma_{gj2} - \sigma_{gj1}} \right)} + D_{gjn\mu} \sin{\left( \frac{n \pi \tau}{\sigma_{gj2} - \sigma_{gj1}} \right)},\\
        T_{gj0\mu}(\tau) &= C_{gj0\mu} + D_{gj0\mu} \tau.
    \end{aligned}
\end{equation}
Then the solution to the problem may be expressed as Fourier series:
\begin{equation*}
    x_{gj\mu}(\tau, \sigma) = T_{gj0\mu}(\tau)u_{gj0}(\sigma) + \sum_{n=1}^{+\infty} T_{gjn\mu}(\tau)u_{gjn}(\sigma).
\end{equation*}
The first initial condition gives expressions for the coefficients $C_{gjn\mu}$:
\begin{equation}
    \label{eq:fragm_D}
    \begin{aligned}
        C_{gjn\mu} & = \frac{2}{\sigma_{gj2} - \sigma_{gj1}} \int_{\sigma_{gj1}}^{\sigma_{gj2}} \cos{\left( \pi n \frac{\sigma - \sigma_{gj1}}{\sigma_{gj2} - \sigma_{gj1}} \right)} \rho_{gj\mu}(\sigma)d\sigma,\\
        C_{gj0\mu} &= \frac{1}{\sigma_{gj2} - \sigma_{gj1}} \int_{\sigma_{gj1}}^{\sigma_{gj2}} \rho_{gj\mu}(\sigma)d\sigma.
    \end{aligned}
\end{equation}
The second condition yields $D_{gjn\mu}$ coefficients:
\begin{equation}
    \label{eq:fragm_C}
    \begin{aligned}
        D_{gjn\mu} & = \frac{2}{\pi n} \int_{\sigma_{gj1}}^{\sigma_{gj2}} \cos{\left( \pi n \frac{\sigma - \sigma_{gj1}}{\sigma_{gj2} - \sigma_{gj1}} \right)} v_{gj\mu}(\sigma)d\sigma,\\
        D_{gj0\mu} &= \frac{1}{\sigma_{gj2} - \sigma_{gj1}} \int_{\sigma_{gj1}}^{\sigma_{gj2}} v_{gj\mu}(\sigma)d\sigma.
    \end{aligned}
\end{equation}
Let us introduce the notations:
\begin{equation}
    \label{eq:fragm_P_def}
    P_{gj\mu} \equiv \kappa \int_{\sigma_{gj1}}^{\sigma_{gj2}} v_{gj\mu}(\sigma)d\sigma,
\end{equation}
\begin{equation}
    \label{eq:fragm_Q_def}
    Q_{gj\mu} \equiv \frac{1}{\sigma_{gj2} - \sigma_{gj1}} \int_{\sigma_{gj1}}^{\sigma_{gj2}} \rho_{gj\mu}(\sigma)d\sigma,
\end{equation}
\begin{equation}
    \label{eq:fragm_famp_def}
    \alpha_{gjn\mu} \equiv \sqrt{\frac{\kappa}{\pi}}  \int_{\sigma_{gj1}}^{\sigma_{gj2}} \cos{\left( \pi n \frac{\sigma - \sigma_{gj1}}{\sigma_{gj2} - \sigma_{gj1}} \right)} \left[ v_{gj\mu}(\sigma) - i\frac{\pi n}{\sigma_{gj2}-\sigma_{gj1}}\rho_{gj\mu}(\sigma) \right] d\sigma.
\end{equation}
Then the solution to the boundary value problem for the motion of the daughter string $j$ of the generation $g$ is given by a formula:
\begin{equation}
    \label{eq:fragm_x_gen}
    \begin{aligned}
         x_{gj\mu}(\tau, \sigma) &= Q_{gj\mu} + \frac{P_{gj\mu}\tau}{\kappa(\sigma_{gj2} - \sigma_{gj1})} \\
         &+ \frac{i}{\sqrt{\kappa \pi}} \sum_{n \ne 0} \frac{\alpha_{gjn\mu}}{n} \exp{\left( - \frac{in\pi\tau}{\sigma_{gj2} - \sigma_{gj1}} \right)} \cos{\left( \pi n \frac{\sigma - \sigma_{gj1}}{\sigma_{gj2} - \sigma_{gj1}} \right)}.
    \end{aligned}
\end{equation}
Note that in the case of extreme values $\sigma_{gj1}=0$, $\sigma_{gj2}=\pi$ formula (\ref{eq:fragm_x_gen}) becomes (\ref{eq:solution_std}). One can question whether it is possible to use the same statement of the Cauchy problem as (\ref{eq:CauchyProb}) for daughter strings. The issue here is that initial data functions must satisfy conservation laws, so if $\sigma$-reparametrization took place when transitioning from mother string to daughters, the initial data functions cannot simply be defined as mother string configuration at the break time. One has to introduce the renormalization factors to modify $\rho_{gj\mu}(\sigma)$, $v_{gj\mu}(\sigma)$. This leads to a complication in understanding the meaning of the quantities we operate. Moreover, such a procedure leads to the same numeric coefficients in the resulting solution as in (\ref{eq:fragm_x_gen}). So, both approaches are equivalent, and the latter will not be considered here in detail.

\subsection{The single-point fragmentation of the FOEE(1)-string}
\label{sec:foee1_fragm}
Let the $1^{\xi}_{\nu}$-string be defined by the FOEE(1) method with mass $M$ and 4-momentum $P_{\mu}$. Its motion is described by formulae (\ref{eq:nustr_x}) - (\ref{eq:nustr_Lambda}) of Section \ref{sec:nustr}. Let the string break occur at the time $\tau = \tau^*$ at the point $\sigma = \sigma^*$. In order to comply with the notation introduced in the previous subsection, we assign generation $g=0$ to these parameters and index $j=1$ to the breaking point (since there is only one): $\tau^* \equiv \tau_{01}^*$, $\sigma^* \equiv \sigma_{01}^*$. Then the initial data functions for problems for the motion of first-generation string fragments are:
\begin{equation}
    \label{eq:fragm_ic_firstgen}
    \begin{aligned}
        \rho_{1j\mu}(\sigma) &\equiv x_{\mu}(\tau_{01}^*,\sigma) = (\kappa \pi)^{-1} \left[ P_{\mu} \tau_{01}^* + \nu^{-1} \left( M\Omega_{\mu}(\tau_{01}^*, \nu) - (P\Omega(\tau_{01}^*,\nu))\chi_{\mu} \right) \cos(\nu\sigma) \right],\\
        v_{1j\mu}(\sigma) &\equiv \dot{x}_{\mu}(\tau_{01}^*,\sigma) = (\kappa \pi)^{-1} \left[ P_{\mu} + \xi\left( M\Lambda_{\mu}(\tau_{01}^*, \nu) - (P\Lambda(\tau_{01}^*,\nu))\chi_{\mu} \right) \cos(\nu\sigma) \right],\\
    \end{aligned}
\end{equation}
$j=1,~2$, where $\sigma \in [0, \sigma_{01}^*]$ for $j=1$ and $\sigma \in [\sigma_{01}^*, \pi]$ for $j=2$, and the first index $g=1$ of initial data functions defines the generation of daughter strings.

The linear terms and expressions before the cosines in (\ref{eq:fragm_ic_firstgen}) are nothing more than coefficients. To simplify the expressions we can introduce the following notation:
\begin{equation}
    \label{eq:fragm_notation_firstgen}
    \begin{aligned}
        a_{1(1,2)\mu} &\equiv (\kappa \pi)^{-1} P_{\mu},\\ 
        b_{1(1,2)\mu} &\equiv (\kappa \pi)^{-1} \xi \left[ M\Lambda{\mu}(\tau_{01}^*,\nu) - (P\Lambda(\tau_{01}^*,\nu))\chi_{\mu} \right], \\
        c_{1(1,2)\mu} &\equiv (\kappa \pi)^{-1} P_{\mu} \tau_{01}^*,\\ 
        d_{1(1,2)\mu} &\equiv \nu^{-1}(\kappa \pi)^{-1} \left[ M\Omega_{\mu}(\tau_{01}^*,\nu) - (P\Omega(\tau_{01}^*,\nu))\chi_{\mu} \right].
    \end{aligned}
\end{equation}
Here, the notation $(1,2)$ in the coefficient indices represents the fact that the same values are used for both daughter strings (with $j=1,~2$). Thus, the initial conditions for daughter strings are:
\begin{equation}
    \label{eq:fragm_ic_firstgen_short}
    \begin{aligned}
        v_{1j\mu}(\sigma) &= a_{1(1,2)\mu} + b_{1(1,2)\mu} \cos(\nu\sigma), \quad \rho_{1j\mu}(\sigma) = c_{1(1,2)\mu} + d_{1(1,2)\mu} \cos(\nu\sigma),\\
        j &= 1,~2; \quad \sigma \in [0, \sigma_{01}^*], \quad j = 1; \quad \sigma \in [\sigma_{01}^*, \pi], \quad j = 2.
    \end{aligned}
\end{equation}
Let us derive the expressions for the Fourier amplitudes. For $j=1$ (by substituting (\ref{eq:fragm_ic_firstgen_short}) in (\ref{eq:fragm_famp_def})) we get:
\begin{equation}
    \label{eq:fragm_famp11}
    \begin{aligned}
        \alpha_{11n\mu} 
        &=\sqrt{\frac{\kappa}{\pi}} \left( \left(a_{1(1,2)\mu} - i\frac{\pi n}{\sigma_{01}^*}c_{1(1,2)\mu}\right) \int_{0}^{\sigma_{01}^*} \cos{\left( \frac{n\pi\sigma}{\sigma_{01}^*} \right)} d\sigma \right.\\
        &\quad\quad\quad+ \left. \left(b_{1(1,2)\mu} - i\frac{\pi n}{\sigma_{01}^*}d_{1(1,2)\mu}\right) \int_{0}^{\sigma_{01}^*} \cos{\left( \frac{n\pi\sigma}{\sigma_{01}^*} \right)}\cos(\nu\sigma)d\sigma \right)\\
        &= \sqrt{\frac{\kappa}{\pi}} \left(b_{1(1,2)\mu} - i\frac{\pi n}{\sigma_{01}^*}d_{1(1,2)\mu}\right) \frac{\nu \cos(\pi n) \sin(\nu\sigma_{01}^*)}{\nu^2-(\pi n/\sigma_{01}^*)^2}, \quad n \ne \nu.
    \end{aligned}
\end{equation}
For $j=2$:
\begin{equation}
    \label{eq:fragm_famp12}
    \begin{aligned}
        \alpha_{12n\mu} = -\sqrt{\frac{\kappa}{\pi}} \left(b_{1(1,2)\mu} - i\frac{\pi n}{\pi-\sigma_{01}^*}d_{1(1,2)\mu}\right) \frac{\nu  \sin(\nu\sigma_{01}^*)}{\nu^2-(\pi n/(\pi - \sigma_{01}^*))^2},\quad n \ne \nu.
    \end{aligned}
\end{equation}
The expressions (\ref{eq:fragm_famp11}), (\ref{eq:fragm_famp12}), are not equal to zero in the case of arbitrary order $n$. This means that the system of Virasoro conditions (\ref{eq:virasoro_cond}) consists of an infinite number of equations with a finite number of coefficients, which makes it unsolvable.

However, there is a way out of this situation. It is clear that if the breaking point $\sigma_{01}^*$ belongs to a special set of points on the string $\sigma_{01}^* = \pi r/\nu$, where $r$ is a natural number, $0<r<\nu$, then expressions (\ref{eq:fragm_famp11}) and (\ref{eq:fragm_famp12}) turn out to be equal to zero. Then the Virasoro system (\ref{eq:virasoro_cond}) turns out to be finite and has the usual form for the FOEE(1) method.

It turns out that a string can only break at a countable set of points (the ``joints''), and the higher the order of the eigenharmonic of the string, the more break points are allowed. In the following, we will refer to the number $r$ as the ``cut factor'' of the string and assign it the same indices as the break point itself: $r_{gj}$ is the cut factor of the $j$-th string of generation $g$. Thus, the string breaks at the point with the coordinate
\begin{equation}
    \label{eq:fragm_break_point}
    \sigma_{gj}^* = \frac{\pi r_{gj}}{\nu}, \quad j = 1,~\ldots,~2^g.
\end{equation}

\begin{figure}[htbp]
    \centering
    \includegraphics[width=.6\textwidth]{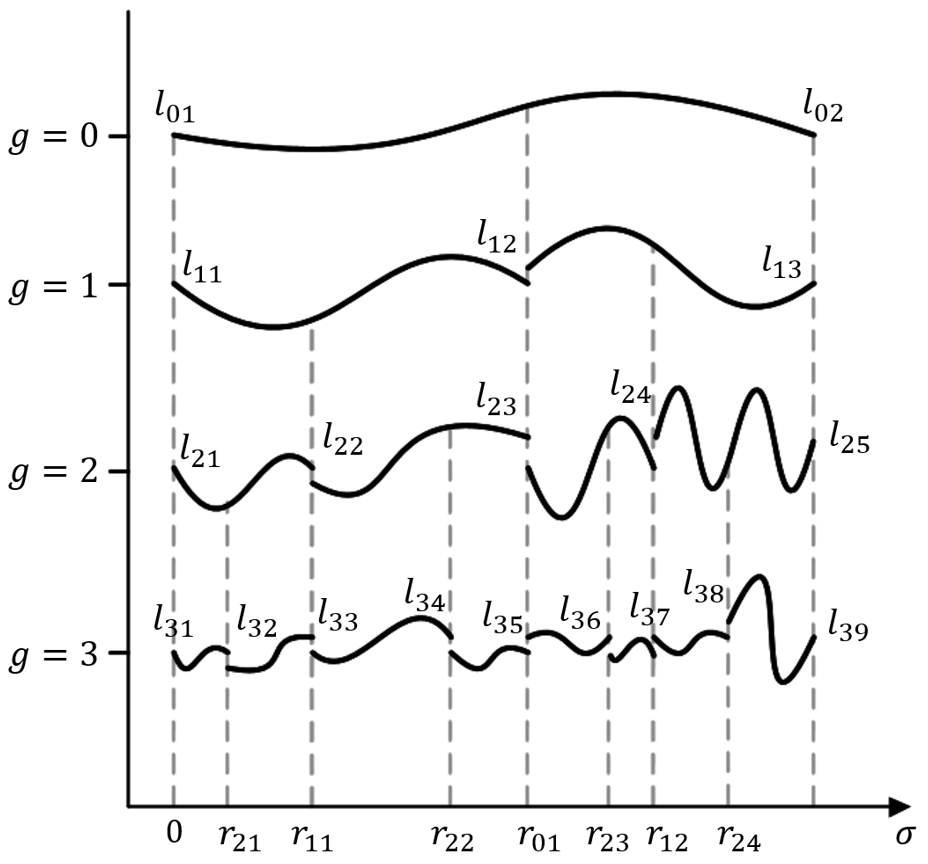}
    \caption{A schematic representation of the string fragmentation process. The generation of daughter strings is marked on the vertical axis. The shape of the different strings is intentionally shown to be very different in order to emphasize their independent existence.\label{fig:fig5}}
\end{figure}

It is necessary to determine the limits within which the value of the cut factor can be sampled for each string in a generation. Several stages of string fragmentation are schematically depicted in figure \ref{fig:fig5}.
The primary string of generation $g=0$ breaks at the point $\sigma_{01}^* = \pi r_{01} / \nu$. The cut factor $r_{01}$ can be sampled within the limits $0 < r_{01} < \nu$, which corresponds to the coordinates $0 < \sigma_{01}^* < \pi$ of the break point. Let us designate the limits of the interval of possible values of the cut factor for the string of the zeroth generation as $l_{01} \equiv 0$, $l_{02} \equiv \nu$. The strings of the first generation break at the points $\sigma_{11}^*$ (the first string from the end $\sigma = 0$, the left end) and $\sigma_{12}^*$ (the second string), respectively. The cut factor of the first string can take integer values in the interval $0 < r_{11} < r_{01}$, and the cut factor of the second one is in the interval $r_{01} < r_{12} < \nu$. The interval limits for the cut factors of the first generation strings are, thus, as follows: $l_{11} \equiv l_{01} = 0$, $l_{12} \equiv r_{01}$, $l_{13} \equiv l_{02} = \nu$. For second generation strings, the cut factors lie in the intervals:
\begin{equation*}
    \begin{aligned}
        l_{21} \equiv 0 &< r_{21} < r_{11} \equiv l_{22}, \quad l_{22} < r_{22} < r_{01} \equiv l_{23},\\
        l_{23} &< r_{23} < r_{12} \equiv l_{24}, \quad l_{24} < r_{24} < \nu \equiv l_{25}.
    \end{aligned}
\end{equation*}
Similarly, one can generalize this procedure to an arbitrary generation of strings. Thus, the cut factors $r_{gj}$, $j = 1,~\ldots,~2^g$ for strings of generation $g$ are sampled in intervals:
\begin{equation}
    \label{eq:l_interv}
    l_{gj} < r_{gj} < l_{g(j+1)}, \quad j = 1,~\ldots,~2^g,
\end{equation}
where the boundaries are determined as
\begin{equation}
    \label{eq:l_bound}
    l_{g(2k-1)} = l_{(g-1)k} , \quad l_{g(2k)} = r_{(g-1)k}, \quad k = 1,~\ldots,~2^{g-1}+1, \quad l_{01} = 0, \quad l_{02} = \nu.
\end{equation}
Taking into account Eqs. (\ref{eq:fragm_w_u}) and (\ref{eq:l_bound}) allows to represent the eigenfunctions as:
\begin{equation}
    \label{eq:fragm_u}
    u_{gjn}(\sigma) = \cos{\left( n\frac{\nu\sigma - \pi l_{gj}}{l_{g(j+1)} - l_{gj}} \right)}, \quad n=1,~2,~\ldots~.
\end{equation}
The orthogonality of eigenfunctions has the form:
\begin{equation}
    \label{eq:fragm_u_ort}
    \int_{\sigma_{gj1}}^{\sigma_{gj2}} u_{gjn}(\sigma)u_{gjm}(\sigma)d\sigma = \frac{l_{g(j+1)}-l_{gj}}{\nu}\frac{\pi}{2}\delta_{nm}.
\end{equation}
The Fourier amplitudes can now be calculated explicitly. Here we will assume that the oscillation of the initial condition functions for the $g$-th generation of strings is determined by a function proportional to $\cos(\nu\sigma)$, just like for the primary string, since it is clear that due to the orthogonality properties (\ref{eq:fragm_u_ort}) only the Fourier amplitudes corresponding to this harmonic should remain non-zero for strings of each generation. Substituting the initial conditions for the string $j$ of generation $g$ in a form, similar to (\ref{eq:fragm_ic_firstgen_short}), and eigenfunctions (\ref{eq:fragm_u}) into (\ref{eq:fragm_famp_def}), as well as taking into account (\ref{eq:fragm_u_ort}) one can obtain the formula:
\begin{equation}
    \label{eq:fragm_famp}
    \alpha_{gjn\mu} = \frac{\sqrt{\kappa\pi}}{2} \left(\frac{l_{g(j+1)} - l_{gj}}{\nu}b_{gj\mu} - ind_{gj\mu} \right) \cos(\pi l_{gj}) \delta_{(l_{g(j+1)} - l_{gj})|n|},
\end{equation}
where $a_{gj}$, $b_{gj}$, $c_{gj}$ and $d_{gj}$ are the coefficients defined similarly to (\ref{eq:fragm_notation_firstgen}) for this string.

To obtain the expressions for the coefficients $a_{gj}$, $b_{gj}$, $c_{gj}$, and $d_{gj}$, one first needs to substitute (\ref{eq:fragm_famp}) into (\ref{eq:fragm_x_gen}):
\begin{equation*}
    \begin{aligned}
        x_{gj\mu}(\tau,\sigma) &= \frac{P_{\mu}\tau}{\kappa \pi}+ c_{gj\mu} \\ 
        &+ \frac{i}{\sqrt{\kappa \pi}} \sum_{n = \pm (l_{g(j+1)} - l_{gj})} \frac{\alpha_{gjn\mu}}{n} \exp{\left( -\frac{in\nu\tau}{l_{g(j+1)} - l_{gj}} \right)} \cos{\left( n \frac{\nu\sigma - \pi l_{gj}}{l_{g(j+1)} - l_{gj}} \right)}\\
        =\frac{P_{\mu}\tau}{\kappa \pi} &+ c_{gj\mu} + \frac{i}{\sqrt{\kappa \pi}} \frac{\alpha_{gj(l_{g(j+1)}-l_{gj})}e^{-i\nu\tau} - \alpha_{gj[-(l_{g(j+1)}-l_{gj})]}e^{i\nu\tau}}{l_{g(j+1)}-l_{gj}} \cos(\nu\sigma - \pi l_{gj}).
    \end{aligned}
\end{equation*}
We can simplify:
\begin{equation*}
    \begin{aligned}
        &\alpha_{gj(l_{g(j+1)}-l_{gj})}e^{-i\nu\tau} - \alpha_{gj[-(l_{g(j+1)}-l_{gj})]}e^{i\nu\tau}\\ 
        &=-i \sqrt{\kappa \pi} \frac{l_{g(j+1)}-l_{gj}}{\nu} \cos(\pi l_{gj}) \left[ b_{gj\mu} \sin(\nu\tau) + \nu d_{gj\mu} \cos(\nu\tau) \right].
    \end{aligned}
\end{equation*}
The formula turns into
\begin{equation}
    \label{eq:fragm_x_bcd}
    \begin{aligned}
        x_{gj\mu}(\tau,\sigma) = \frac{P_{\mu}\tau}{\kappa \pi}+ c_{gj\mu} 
        +  \left[\nu^{-1} b_{gj\mu} \sin(\nu\tau) + d_{gj\mu} \cos(\nu\tau) \right] \cos(\nu\sigma).
    \end{aligned}
\end{equation}
Using (\ref{eq:fragm_x_bcd}), one can obtain the recurrent formulae for the parameters $a_{gj\mu}$, $b_{gj\mu}$, $c_{gj\mu}$ and $d_{gj\mu}$:
\begin{equation}
    \label{eq:fragm_a_rec}
    a_{(g+1)(2j-1,2j)\mu} = \frac{P_{\mu}}{\kappa \pi} \equiv a_{\mu},
\end{equation}
\begin{equation}
    \label{eq:fragm_b_rec}
    b_{(g+1)(2j-1,2j)\mu} = b_{gj\mu} \cos(\nu\tau_{gj}^*) - \nu d_{gj\mu} \sin(\nu\tau_{gj}^*),
\end{equation}
\begin{equation}
    \label{eq:fragm_c_rec}
    c_{(g+1)(2j-1,2j)\mu} = \frac{P_{\mu} \tau_{gj}^*}{\kappa \pi} + c_{gj\mu},
\end{equation}
\begin{equation}
    \label{eq:fragm_d_rec}
    d_{(g+1)(2j-1,2j)\mu} = \nu^{-1} b_{gj\mu} \sin(\nu\tau_{gj}^*) + d_{gj\mu} \cos(\nu\tau_{gj}^*).
\end{equation}
In Eqs. (\ref{eq:fragm_a_rec}) - (\ref{eq:fragm_d_rec}) $j=1,~\ldots,~2^g$, notation $(2j-1,2j)$ symbolizes that both daughter strings of the same mother have the same coefficients in their initial conditions. Using formula (\ref{eq:fragm_x_bcd}) with (\ref{eq:fragm_a_rec}) - (\ref{eq:fragm_d_rec}), (\ref{eq:l_bound}) and (\ref{eq:fragm_notation_firstgen}) allows one to calculate the motion of the daughter string $j$ in arbitrary generation $g$.

The total momentum of the daughter string is a fraction of the primary string momentum:
\begin{equation}
    \label{eq:fragm_tot_mom}
    P_{gj\mu} = \frac{l_{g(j+1)} - l_{gj}}{\nu} P_{\mu},
\end{equation}
which also produces the same relation for the masses of daughter strings:
\begin{equation}
    \label{eq:fragm_mass}
    M_{gj} = \frac{l_{g(j+1)} - l_{gj}}{\nu} M.
\end{equation}
Note that formula (\ref{eq:fragm_mass}) forbids the existence of a fragment of the primary string that is lighter than $M/\nu$. This result has huge importance as it demonstrates the appearance of the natural limit to the string fragmentation process, something that is absent in regular string fragmentation models and is fixed manually.

To further develop this idea, one can adopt a model in which the upper limit for the parameter $\nu$ is set by considering that the value $M/\nu$ must be of the order of the allowed lightest particle mass, e.g., the pion mass $m_{\pi}$. That would mean that a string with $M=100$ GeV would have a maximum eigenharmonic order of $\nu_{\text{max}}\approx 700$. However, such a mechanism for constraining the upper limit of the parameter $\nu$ could be disputed. 

\subsection{Virasoro conditions and conservation laws for the daughter strings}
\label{sec:virasoro_daughter}
It is important to study the question of Virasoro conditions for the daughter strings. As string breaking is not predefined in the initial Nambu-Goto theory, restrictions imposed by the Virasoro conditions may not be fulfilled for daughter strings, even though all other quantities are conserved.

Fourier amplitudes (\ref{eq:fragm_famp}) yield the following system when substituted in Virasoro conditions (\ref{eq:virasoro_cond}):
\begin{equation}
    \label{eq:fragm_virasoro}
    \left\{
    \begin{aligned}
        b_{gj}P = d_{gj}P = b_{gj}d_{gj} &= 0\\
        b_{gj}^2 - \nu^2d_{gj}^2 &= 0\\
        b_{gj}^2 + \frac{M^2}{(\kappa \pi)^2} &= 0.
    \end{aligned}
    \right.
\end{equation}
Here $P_{\mu}$ is the total momentum of the primary string that initiated the string decay sequence, $M$ is its mass. Note that the second equation has the same factor $\nu^2$ before $d_{gj}^2$, as for the primary string.

To prove that the coefficients defined in (\ref{eq:fragm_a_rec}) - (\ref{eq:fragm_d_rec}) indeed satisfy the system (\ref{eq:fragm_virasoro}), we will use the induction method. Let the system of Virasoro conditions (\ref{eq:fragm_virasoro}) be fulfilled for the coefficients of initial data functions of the string number $j$ of generation $g$. We know that such a string exists, as in the case of a primary $1^\xi_\nu$-string with $g=0$ the Virsoro conditions are certainly satisfied. Then let us see, if the coefficients of the initial data functions of the daughters of this string satisfy the Virasoro conditions. One can check (here the indices that numerate the string in its generation will be omitted, as we only consider the cases of the mother and its two daughters):
\begin{equation*}
    \begin{aligned}
        b_{g+1}P &= (b_{g}P) \cos(\nu\tau_{g}^*) - \nu (d_{g}P) \sin(\nu\tau_{g}^*) \equiv 0,\\
        d_{g+1}P &= \nu^{-1} (b_{g}P) \sin(\nu\tau_{g}^*) + (d_{g}P) \cos(\nu\tau_{g}^*) \equiv 0,\\
        b_{g+1}d_{g+1} &= \left[b_{g\mu} \cos(\nu\tau_{g}^*) - \nu d_{g\mu} \sin(\nu\tau_{g}^*) \right] \left[ \nu^{-1} b_{g}^\mu \sin(\nu\tau_{g}^*) + d_{g}^\mu \cos(\nu\tau_{g}^*) \right] \\
        &= \nu^{-1} \left( b_{g}^2 - \nu^2 d_{g}^2\right)  \cos(\nu\tau_{g}^*) \sin(\nu\tau_{g}^*) \equiv 0,\\
        b_{g+1}^2 - \nu^2 d_{g+1}^2 &= \left[ b_{g\mu} \cos(\nu\tau_{g}^*) - \nu d_{g\mu} \sin(\nu\tau_{g}^*) \right]^2 - \nu^2 \left[ \nu^{-1} b_{g\mu} \sin(\nu\tau_{g}^*) + d_{g\mu} \cos(\nu\tau_{g}^*) \right]^2\\
        &= b_{g}^2 \cos^2(\nu\tau_g^*) + \nu^2 d_g^2 \sin^2(\nu\tau_g^*) - b_g^2 \sin^2(\nu\tau_g^*) - \nu^2 d_g^2 \cos^2(\nu\tau_g^*)\\
        & = \left( b_{g}^2 - \nu^2 d_g^2 \right) \left( \cos^2(\nu\tau_g^*) - \sin^2(\nu\tau_g^*) \right) \equiv 0,\\
     \end{aligned}
\end{equation*}
\begin{equation*}
    \begin{aligned}
        b_{g+1}^2 + \frac{M^2}{(\kappa \pi)^2} &= \left[ b_{g\mu} \cos(\nu\tau_{g}^*) - \nu d_{g\mu} \sin(\nu\tau_{g}^*) \right]^2 + \frac{M^2}{(\kappa \pi)^2} =\\
        & =  b_{g}^2 \cos^2(\nu\tau_g^*) + \nu^2 d_g^2 \sin^2(\nu\tau_g^*) + \frac{M^2}{(\kappa \pi)^2} = b_{g}^2 + \frac{M^2}{(\kappa \pi)^2} \equiv 0.
    \end{aligned}
\end{equation*}
One can see that, if the Virasoro conditions (\ref{eq:fragm_virasoro}) are satisfied for some string of generation $g$, then they are also satisfied for its daughters of generation $g+1$. This implies that for any string of arbitrary generation $G$, $G>g$, this procedure may be repeated. So, as this chain can always start with the primary string according to the induction method, the statement that the system (\ref{eq:fragm_virasoro}) is satisfied for all daughter strings is now proved.

The conservation of total momentum after the string fragmentation is obvious from the definition (\ref{eq:fragm_a_rec}) and the orthogonality condition (\ref{eq:fragm_u_ort}). But the case of total angular momentum is worth looking at. By the definition
\begin{equation*}
    \begin{aligned}
        \mathcal{M}_{gj\mu\lambda} &= \kappa \int_{\sigma_{gj1}}^{\sigma_{gj2}} d\sigma \left[ \left( c_{gj\mu} + d_{gj\mu} \cos(\nu\sigma) \right) \left( a_{gj\lambda} + b_{gj\lambda} \cos(\nu \sigma) \right) - (\mu \leftrightarrow \lambda) \right]\\
        &= \frac{l_{g(j+1)} - l_{gj}}{\nu} \pi \left[ c_{gj\mu} a_{gj\lambda} - c_{gj\lambda}a_{gj\mu} + \frac{1}{2} \left( d_{gj\mu} b_{gj\lambda} - d_{gj\lambda} b_{gj\mu} \right) \right],
    \end{aligned}
\end{equation*}
where we can substitute the expression for $a_{gj\mu}$ to obtain the formula for the total angular momentum tensor of the daughter string:
\begin{equation}
    \label{eq:fragm_ang_mom}
    \mathcal{M}_{gj\mu\lambda} = \frac{l_{g(j+1)} - l_{gj}}{\nu} \left[ c_{gj\mu} P_{\lambda} - c_{gj\lambda}P_{\mu} + \frac{\kappa \pi}{2} \left( d_{gj\mu} b_{gj\lambda} - d_{gj\lambda} b_{gj\mu} \right) \right].
\end{equation}
The question is whether the following equality is true for any $g$, $j$, $j = 1,~\ldots,~2^g$:
\begin{equation*}
    \mathcal{M}_{(g+1)(2j)\mu\lambda} + \mathcal{M}_{(g+1)(2j+1)\mu\lambda} = \mathcal{M}_{gj\mu\lambda}.
\end{equation*}
It is quite easy to show that:
\begin{equation*}
    \begin{aligned}
        &c_{(g+1)\mu}P_{\lambda} - c_{(g+1)\lambda}P_{\mu} \\
        &= \left( \frac{P_{\mu} \tau_{g}^*}{\kappa \pi} + c_{g\mu} \right)P_{\lambda} - \left( \frac{P_{\lambda} \tau_{g}^*}{\kappa \pi} + c_{g\lambda} \right)P_{\mu}= c_{g\mu}P_{\lambda} - c_{g\lambda}P_{\mu},\\
        &d_{(g+1)\mu}b_{(g+1)\lambda} - d_{(g+1)\lambda}b_{(g+1)\mu} \\
        &= \left( \nu^{-1} b_{g\mu} \sin(\nu\tau_{g}^*) + d_{g\mu} \cos(\nu\tau_{g}^*) \right) \left( b_{g\lambda} \cos(\nu\tau_{g}^*) - \nu d_{g\lambda} \sin(\nu\tau_{g}^*) \right) - (\mu \leftrightarrow \lambda) \\
        &= d_{g\mu}b_{g\lambda} \cos^2(\nu\tau_g^*) - d_{g\lambda}b_{g\mu} \sin^2(\nu\tau_g^*) -(\mu \leftrightarrow \lambda) = d_{g\mu}b_{g\lambda} - d_{g\lambda}b_{g\mu},
    \end{aligned}
\end{equation*}
where the index numerating the string in the generation is omitted, as only the mother and its daughter strings are considered. This means that the total angular momentum of the daughter string is simply a fraction of that of its mother string:
\begin{equation}
    \label{eq:fragm_ang_mom_frac}
    \mathcal{M}_{(g+1)(2j, 2j+1)\mu\lambda} = \frac{l_{(g+1)[(2j,2j+1)+1]} - l_{(g+1)(2j,2j+1)}}{l_{g(j+1)} - l_{gj}} \mathcal{M}_{gj\mu\lambda}, \quad j = 1,~\ldots,~2^g.
\end{equation}
Here the notation $(2j,2j+1)$ once again denotes the applicability of the formula for both string fragments that are enumerated as $2j$ and $2j+1$ in the generation $g+1$. The Eq. (\ref{eq:fragm_ang_mom_frac}) also clearly proves the conservation of the total angular momentum, as after the summation over two daughters the mother angular momentum tensor will be preceded by the fraction $[l_{(g+1)(2j+2)}-l_{(g+1)(2j)}]/[l_{g(j+1)}-l_{gj}]$, which is (following the notation (\ref{eq:l_bound})) just unity. One can rewrite the formula (\ref{eq:fragm_ang_mom_frac}) as
\begin{equation}
    \label{eq:fragm_ang_mom_fractot}
    \mathcal{M}_{gj\mu\lambda} = \frac{l_{g(j+1)} - l_{gj}}{\nu} \mathcal{M}_{\mu\lambda} \equiv \frac{\sigma_{gj2} - \sigma_{gj1}}{\pi} \mathcal{M}_{\mu\lambda}, \quad j = 1,~\ldots,~2^g, \quad g = 1,~2,~\ldots,
\end{equation}
to express the angular momentum of arbitrary daughter string using the angular momentum tensor of the primary string. The equation (\ref{eq:fragm_ang_mom_fractot}) means that the angular momentum component of the string produced as a result of the fragmentation process cannot be less than $\mathcal{M}_{\mu\lambda}/\nu$. In the center-of-mass frame this assures that for a given primary string with spin $J$ the fragments of this string will always have spin not less then $J/\nu$. Thus, if the order $\nu$ of the eigenoscillation of the string with large spin is small, its fragments must also have considerable angular momentum.

The formula (\ref{eq:fragm_ang_mom_fractot}) along with expressions (\ref{eq:fragm_mass}) and (\ref{eq:nustr_spinmass}) produce the relation between the mass of the daughter string and its spin:
\begin{equation}
    \label{eq:fragm_mass_spin}
    J_{gj} = \frac{1}{l_{g(j+1)} - l_{gj}} \frac{M_{gj}^2}{2\kappa \pi}, \quad j = 1,~\ldots,~2^g, \quad g = 1,~2,~\ldots~.
\end{equation}
The formula (\ref{eq:fragm_mass_spin}) means that the applied scheme of fragmentation leads to the property of lighter strings to approach the universal limit of the string spin-mass relation (\ref{eq:spin_limit}), which is often used to draw the connection between the string model and the Regge trajectories \cite{Rebbi, Lizzi, barbashov_nesterenko, Leopoldo}. In this way, the proposed FOEE model predicts the close-to-Regge behavior of the light string fragments (with a steeper slope of the trajectories) regardless of the much flatter slope (\ref{eq:nustr_spinmass}) of the large-mass primary strings.

\subsection{The fragmentation of the FOEE(1)-string with energy release}
\label{sec:fragm_with_gap}
The fragmentation scheme developed in Subsections \ref{sec:foee1_fragm} and \ref{sec:virasoro_daughter} describes a single-point string breaking, i.e. the neighboring endpoints of the new strings are designated with the same $\sigma \equiv \sigma^*$ coordinate. However, there is a significant drawback in this approach. The issue arises when considering the fragmentation of a string in its center-of-mass system. This configuration can model, for example, the two-parton system in $e^+e^-$ collisions. So, the resulting particle spectra predicted by the model must be able to describe the experimental data. But the use of the single-point fragmentation scheme will obviously give an unrealistic picture: according to Eq. (\ref{eq:fragm_tot_mom}) all daughter strings of any generation are produced at rest. Thus, we obtain the scheme for multiparticle production with no energy release. This can be clearly seen from the expression for the mass of the daughter string (\ref{eq:fragm_mass}): the sum of the masses of the daughters gives the mass of the mother string, which is equal to the total energy of the system in the case of the string at rest. One might find this counterintuitive, as our regular perception is that when the stretched string breaks, the new endpoints of the two pieces gain momentum along the string line. However, let me remind that, although the coefficient $\kappa$ in action (\ref{eq:NGaction}) is usually referred to as the ``string tension'' and the examples of the Nambu-Goto string motion often demonstrate the elastic properties of the string, in reality the Nambu-Goto action considers the string with no inner forces of tension (see, e.g., the derivation of the string action in ref. \cite{NGact_3}).

There is no doubt that to obtain the realistic process of string breaking, the proposed scheme should be modified to allow the energy release with two strings gaining additional momentum (and angular momentum). From the point of view of the integral string properties, this should be considered as a two-body decay, but we also need to develop a non-controversial way to describe this using the language of the distributed-along-string quantities. One can start the definition of the initial conditions for the daughter strings completely from scratch using the algorithm defined in the Subsection \ref{sec:alg} for given masses of daughter strings. The downside is that this method basically nullifies the essence of string theory, as it only considers the quantities defining some compound body with mass and momentum. The break point, the configuration of the string, and the way it moves are left out of interest with this approach. Thus, some sort of compromise should be used instead.

Let us consider the case where the string breaks not by discontinuity in a single point $\sigma \equiv \sigma^*$, but by the disappearance of a continuous set of points, $\sigma \in [\sigma^*_1,\sigma^*_2]$. The rest two pieces of the string are the secondary strings (daughters), while the properties of the cut-out chunk of the string redistribute between them. Thus, the daughter strings turn out to be separated by a gap in the $\sigma$-interval.

The basic principles remain quite similar to those developed in Subsections \ref{sec:foee1_fragm}, \ref{sec:virasoro_daughter}. To define the $\sigma$-gap (let us call it $S_{\sigma}$) for each decaying string number $j$ in generation $g$ two cut factors should be sampled now: $S_{\sigma gj} = \{\sigma, ~\sigma \in [\sigma_{g(2j-1)}^*,\sigma_{g(2j)}^*]\}$, where
\begin{equation}
    \label{eq:sigma_gap}
    \begin{aligned}
        \sigma_{g(2j-1)}^* &\equiv r_{g(2j-1)} \pi / \nu, \quad \sigma_{g(2j)}^* \equiv r_{g(2j)} \pi / \nu,\\
        l_{g(2j-1)} < r_{g(2j-1)} &< r_{g(2j)} < l_{g(2j)}, \quad j = 1,~\ldots,~2^g.
    \end{aligned}
\end{equation}
The boundaries for sampling the cut factors for each generation of strings, $g \ge 1$, are determined as
\begin{equation}
    \label{eq:l_bound_gap}
    \begin{aligned}
        l_{g(4j-3)} = l_{(g-1)j}, \quad l_{g(4j-2)} &= r_{(g-1)j}, \quad l_{g(4j-1)} = r_{(g-1)(j+1)}, \quad l_{g(4j)} = l_{(g-1)(j+1)},\\
        j &= 1, ~\ldots, ~2^{g-1}, \quad l_{01} = 0, \quad l_{02} = \nu.
    \end{aligned}
\end{equation}

The first step to define the properties of the string fragments after the vanishing of the string chunk between them would be to conserve the total angular momentum of the system. The reasons for this will become clear later. There are three ways of conserving the angular momentum: changing the rotation speed of the strings, distancing their centers of mass, and increasing their length (mass). The second one seems unnatural, as there is no reason for the daughters to move apart and then change the direction of their momenta (otherwise, no angular momentum will appear). The first one seems plausible, but in fact is also forbidden. The issue lays once again in the Virasoro conditions for the Fourier amplitudes of the daughter strings, as changing the eigenfrequency $\nu \rightarrow \nu^{\prime}$ would cause the terms with sine to not vanish in the expressions, as in the formulae (\ref{eq:fragm_famp11}) and (\ref{eq:fragm_famp12}). The only possible choice left is to modify the masses of the fragments to match the needed angular momentum.

According to (\ref{eq:fragm_ang_mom_frac}), the angular momentum of the missing chunk of the string is
\begin{equation}
    \label{eq:angmom_miss}
    \mathcal{M}_{gj\mu\lambda}^{\text{miss}} = \frac{r_{g(2j)} - r_{g(2j-1)}}{l_{g(2j)} - l_{g(2j-1)}} \mathcal{M}_{gj\mu\lambda}, \quad j = 1,~\ldots,~2^g.
\end{equation}
Suppose that the missing angular momentum spreads evenly between the two fragments. Then, the resulting angular momentum tensors of the daughter strings are:
\begin{equation}
    \label{eq:angmom_mod}
    \begin{aligned}
        \mathcal{M}_{(g+1)(2j-1)\mu\lambda} &= \frac{r_{g(2j)} + r_{g(2j-1)} - 2l_{g(2j-1)}}{2(l_{g(2j)} - l_{g(2j-1)})} \mathcal{M}_{gj\mu\lambda},\\
        \mathcal{M}_{(g+1)(2j)\mu\lambda} &= \frac{2l_{g(2j)} - r_{g(2j)} - r_{g(2j-1)}}{2(l_{g(2j)} - l_{g(2j-1)})} \mathcal{M}_{gj\mu\lambda},\\
    \end{aligned}
\end{equation}
where $j = 1,~\ldots,~2^g$. Note that in this case the simple relation between the spin (and mass) of the daughter string of arbitrary generation and the spin (mass) of the primary string is no longer valid, so formulae (\ref{eq:fragm_tot_mom}), (\ref{eq:fragm_mass}) and (\ref{eq:fragm_ang_mom_fractot}) cannot be used here! The relation between the mass of the daughter string and its spin is still determined by the expression (\ref{eq:fragm_mass_spin}), since it only depends on the eigenvalue $\nu$, the limits of the $\sigma$-interval and the string mass. This allows to calculate the masses of the daughter strings:
\begin{equation}
    \label{eq:fragm_mass_mod}
    \begin{aligned}
        M_{(g+1)(2j-1)} &=   \frac{\sqrt{\left( r_{g(2j-1)} - l_{g(2j-1)} \right) \left( [r_{g(2j)} + r_{g(2j-1)}]/2 - l_{g(2j-1)} \right)}}{l_{g(2j)} - l_{g(2j-1)}}  M_{gj},\\
        M_{(g+1)(2j)} &=   \frac{\sqrt{\left( l_{g(2j)}  - r_{g(2j)} \right) \left( l_{g(2j)} - [r_{g(2j)} + r_{g(2j-1)}]/2 \right)}}{l_{g(2j)} - l_{g(2j-1)}}  M_{gj}.\\
    \end{aligned}
\end{equation}
It is not hard to show that for any given $a < b < c < d$
\begin{equation*}
    \frac{\sqrt{(b-a) \left(\frac{b+c}{2}-a \right)} + \sqrt{(d-c) \left(d - \frac{b+c}{2} \right)}}{d-a} < 1,
\end{equation*}
though the actual proof is too bulky to be given here. Thus, the sum of masses of the daughter strings is less than the mass of the mother string. This means that there will always be a release of energy in the decay of the string.

To make the formulae below more readable, let us omit the $g$, $j$ indices; the mass of the mother string will be $M$, the daughters have masses $M_1$ and $M_2$. Using Eq. (\ref{eq:fragm_mass_mod}) one can always easily calculate the mass of any string considered.

\begin{figure}[htbp]
    \centering
    \begin{subfigure}[b]{0.3\textwidth}
         \centering
         \includegraphics[width=\textwidth]{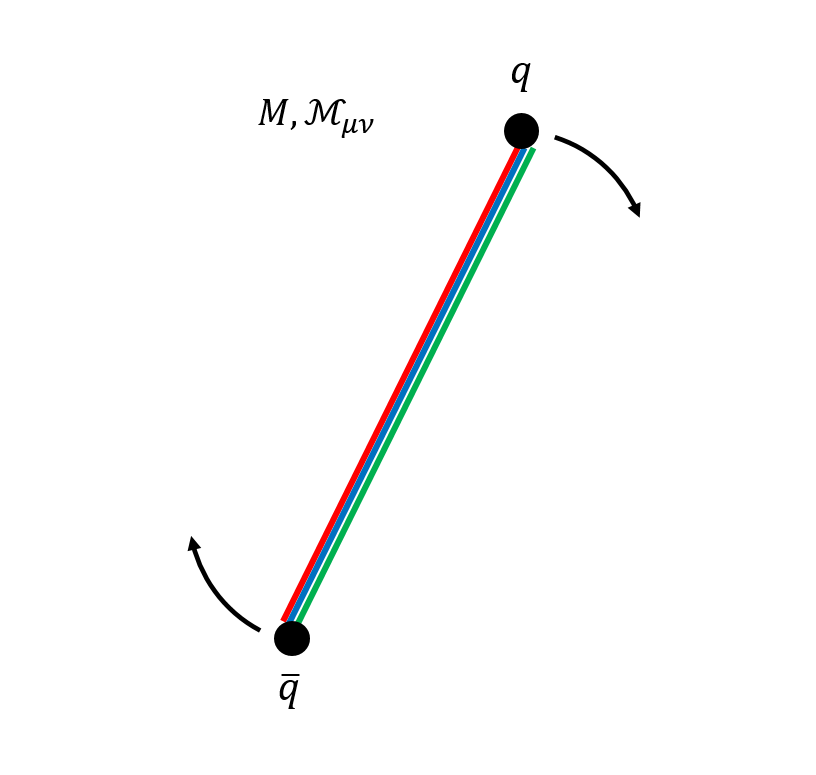}
         \caption{}
         \label{fig:frag_sch_a}
    \end{subfigure}
    \hfill
    \begin{subfigure}[b]{0.3\textwidth}
         \centering
         \includegraphics[width=\textwidth]{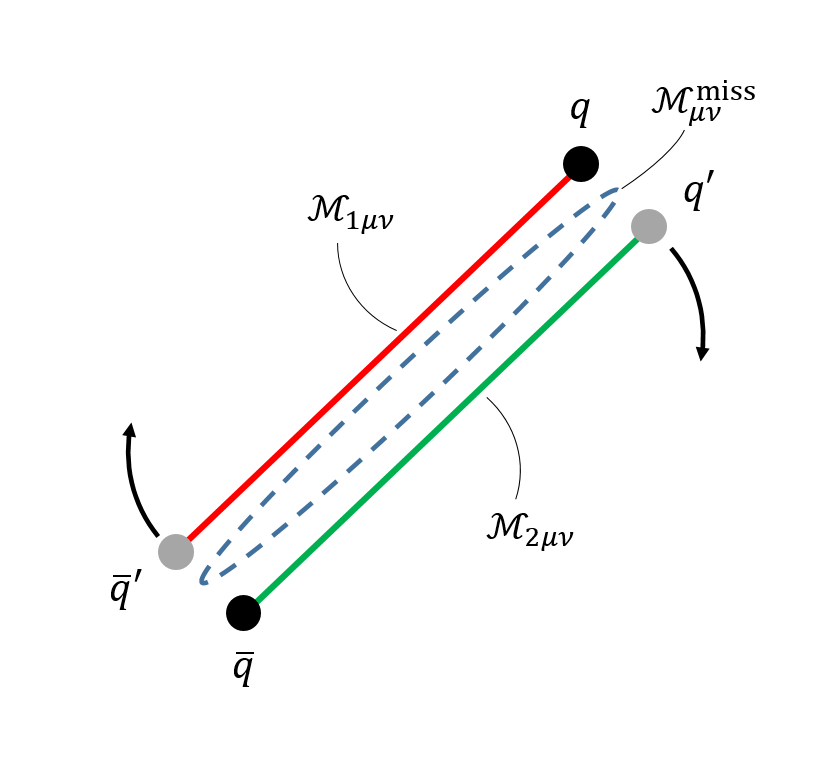}
         \caption{}
         \label{fig:frag_sch_b}
    \end{subfigure}
    \hfill
    \begin{subfigure}[b]{0.3\textwidth}
         \centering
         \includegraphics[width=\textwidth]{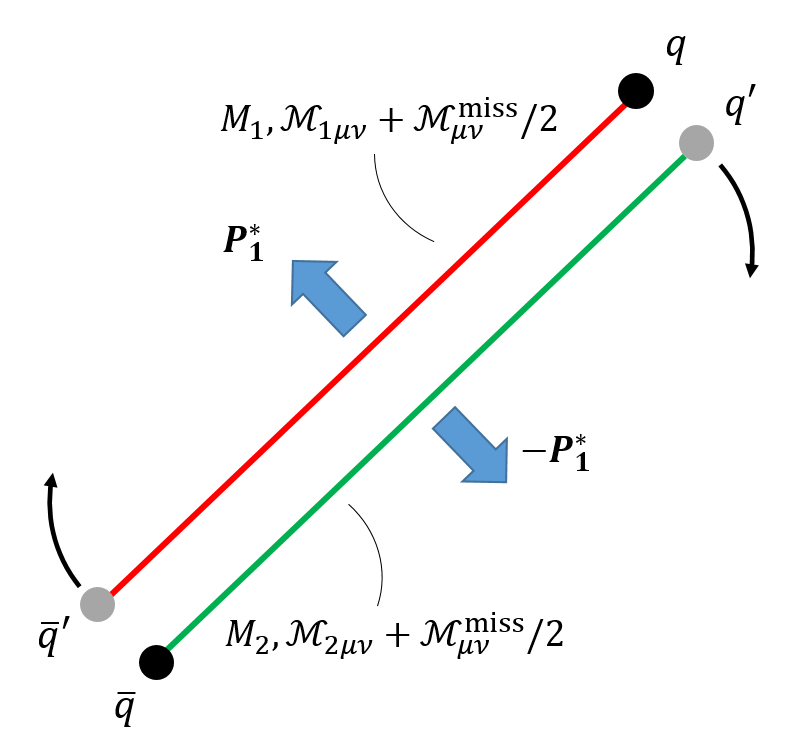}
         \caption{}
         \label{fig:frag_sch_c}
    \end{subfigure}
    \caption{A schematic view of the $1_3^+$-string fragmentation. A ``folded'' string rotates as a rigid rod (figure \ref{fig:frag_sch_a}). Then the string fragments by having the inner section disappear (figure \ref{fig:frag_sch_b}). The missing angular momentum is redistributed between the daughters, their mass is increased, and the released energy is transferred to them by the momentum boost (figure \ref{fig:frag_sch_c}). \label{fig:fragm_scheme}}
\end{figure}

To determine the redistribution of the 4-momentum of the vanished chunk of the string, we will use the two-body decay law. First, the transition to the center-of-mass frame of the mother string should be performed. The total energies of daughter strings $P_1^{*0}$, $P_2^{*0}$ are then determined:
\begin{equation}
    \label{eq:str_energy_cm}
    P_1^{*0} = \frac{M^2+M_1^2-M_2^2}{2M}, \quad P_2^{*0} = \frac{M^2+M_2^2-M_1^2}{2M}.
\end{equation}
The absolute value of their 3-momentum is also known:
\begin{equation}
    \label{eq:str_mom_cm}
    |\boldsymbol{P}^*_{1,2}| = (2M)^{-1}\sqrt{\left( M^2 - \left(M_1 + M_2\right)^2 \right) \left( M^2 - \left(M_1 - M_2\right)^2 \right)}.
\end{equation}
The boost of the daughter strings should be performed in the following way. Let us consider the first daughter string. In the current frame of reference its coordinates and velocity are determined only by the mass of the string and the moment in time $\tau$ (this again follows from the fact that in the center-of-mass frame of the mother string the daughters have zero total momentum 3-vector). This means that one only needs to perform the Lorentz boost to the system where this daughter string has the total momentum $P_{\mu} \equiv (P_1^{*0},\boldsymbol{P_1^*})$ calculated in (\ref{eq:str_energy_cm}), (\ref{eq:str_mom_cm}), using the transformations (\ref{eq:lorentz_boost_string}) and to declare that the daughter string still remains in the center-of-mass frame of the mother string. Let us define the direction of the boost momentum $\boldsymbol{P_1^*}$ to be along the perpendicular to the string line in the rotation plane. Thus, a schematic picture of the string fragmentation turns out to be as shown in figure \ref{fig:fragm_scheme}.

The initial conditions for the daughters can be defined in the following way. We know that any string determined on the $\sigma$-interval $\sigma \in [\sigma_1, \sigma_2]$ at given moment in time $\tau^*$ can be defined in its center-of-mass frame by the expression:
\begin{equation}
    \label{eq:ic_mod}
    x_{\mu}(\tau^*, \sigma) = \frac{M}{\kappa (\sigma_2 - \sigma_1)} \left[ \delta_{0\mu}\tau^* + \frac{ \xi \delta_{3\mu} \sin(\phi+\nu\tau^*) + \delta_{1\mu}\cos(\phi+\nu\tau^*)}{\nu} \cos(\nu \sigma) \right],
\end{equation}
where $M$ is the mass of the string and $\phi$ is the angle that defines the initial rotation of the string. Let the masses of two sections of the string, $\sigma \in S_1$ and $\sigma \in S_2$, (denoted with red and green colors in figure \ref{fig:fragm_scheme}) before the energy redistribution be $M_1^{\text{old}}$ and $M_2^{\text{old}}$, respectively. One only needs to substitute $M_{1,2}^{\text{old}}$ in formula (\ref{eq:ic_mod}) by $M_{1,2}$ defined in (\ref{eq:fragm_mass_mod}). After that, the momentum boost (\ref{eq:str_energy_cm}), (\ref{eq:str_mom_cm}) is applied to account for the energy release. The initial velocity of the daughters is simply the $\tau$-derivative of (\ref{eq:ic_mod}) calculated at $\tau=\tau^*$. Thus, the initial conditions for the daughter strings of the generation $g+1$ in the rest frame of their mother (generation $g$) are defined as:
\begin{equation}
    \label{eq:ic_v_rho_mod}
    \begin{aligned}
        \rho_{(g+1)k_j}^{\mu}(\sigma) &= \beta_{gk_j} \left( \delta^{0\mu}\tau^*_{gj} + \left[ \xi \delta^{3\mu} \sin(\phi+\nu\tau^*_{gj}) + \delta^{1\mu}\cos(\phi+\nu\tau^*_{gj}) \right] \frac{\cos(\nu \sigma)}{\nu} \right),\\
        v_{(g+1)k_j}^{\mu}(\sigma) &= \beta_{gk_j}  \left[ \delta^{0\mu} + \left[ \xi \delta^{3\mu} \cos(\phi+\nu\tau^*_{gj}) - \delta^{1\mu}\sin(\phi+\nu\tau^*_{gj}) \right] \cos(\nu \sigma) \right], \\
        k_j &= 2j-1, ~2j; \quad j = 1, ~\ldots, ~ 2^{g},
    \end{aligned}
\end{equation}
where
\begin{equation}
    \label{eq:beta}
    \beta_{gk_j} = \frac{\nu M_{(g+1)k_j}}{\kappa \pi (l_{(g+1)(2k_j)} - l_{(g+1)(2k_j-1)})}.
\end{equation}

We will not repeat the actions performed previously to obtain the formulae for the Fourier amplitudes, string coordinates, and velocity. One can use the equations (\ref{eq:fragm_x_bcd})-(\ref{eq:fragm_d_rec}) to calculate the motion of the string and (\ref{eq:fragm_famp}) to obtain the amplitudes; only the limits for $\sigma$ now follow the notation (\ref{eq:sigma_gap}), (\ref{eq:l_bound_gap}). Since the rotation frequency of the strings of any generation remains unchanged in the defined fragmentation scheme, we can define the rotation angle of the $j$-th string of generation $G$ as:
\begin{equation}
    \label{eq:rot_ang}
    \phi_{Gj}(\tau) = \sum_{g=1}^{G-1} \phi_{gj_g}^* + \xi\nu \tau \equiv \xi\nu \left(\sum_{g=1}^{G-1}\tau^*_{gj_g} + \tau \right),
\end{equation}
where index $j_g$ denotes the ordinal number of the predecessor string of generation $g$, which initiated the decay sequence that ended up with the considered daughter string. $\phi_{gj_g}^*$ are the rotation angles of the predecessor strings at their break times $\tau^*_{gj_g}$. Thus, we just need to keep the value of the ``global time'' for each string to measure the total time interval since the beginning of the motion of the primary string. The similar simple relation may be obtained for the initial coordinates of the center of mass of the string:
\begin{equation}
    \label{eq:Q_gj}
    Q_{Gj\mu} = \frac{\nu}{\kappa \pi} \sum_{g=1}^{G-1} \frac{P_{gj_g\mu} \tau^*_{gj_g}}{l_{g(2j_g)} - l_{g(2j_g-1)}}.
\end{equation}
The resulting formula for the coordinates of the string is then:
\begin{equation}
    \label{eq:x_final}
    x_{Gj\mu}(\tau, \sigma) = Q_{Gj\mu} + \frac{\nu P_{Gj\mu} \tau + \left[ M_{Gj}\Omega_{Gj\mu}(\tau, \nu) - (P_{Gj}\Omega_{Gj}(\tau,\nu))\chi_{Gj\mu} \right] \cos(\nu\sigma)}{\kappa \pi (l_{g(2j)} - l_{g(2j-1)})},
\end{equation}
where
\begin{equation}
    \label{eq:where}
    \begin{aligned}
        \Omega_{Gj\mu}(\tau, \nu) &\equiv
        \begin{pmatrix}
            0 \\ \cos\varphi \\ \sin\varphi \\ 0
        \end{pmatrix} \cos(\phi_{Gj}(\tau)) + \xi
        \begin{pmatrix}
            0 \\ \sin\varphi\sin\theta \\ -\cos\varphi\sin\theta \\ \cos\theta
        \end{pmatrix}
        \sin(\phi_{Gj}(\tau)),\\
        \chi_{Gj}^0 &\equiv 1, \quad \boldsymbol{\chi}_{Gj} = \frac{\boldsymbol{P}_{Gj}}{P_{Gj}^0 + M_{Gj}},
    \end{aligned}
\end{equation}
$\theta$ and $\varphi$ are the angles defining the orientation of the rotation plane of the string in the chosen coordinate system.

The area decay probability for the daughter strings is only a slightly modified version of (\ref{eq:prob_to_not_decay}):
\begin{equation}
    \label{eq:prtndecay_fragm}
    P_{gj}(\tau) \propto \exp{\left( -\frac{\nu P_0M^2_{gj}\tau}{2\kappa \pi (l_{g(2j)} - l_{g(2j-1)})} \right)}, \quad j = 1,~\ldots,~2^g, \quad g \ge 0.
\end{equation}

\subsection{The algorithm to calculate string fragmentation}
\label{sec:alg_fragm}
The resulting algorithm to compute the iterative process of string fragmentation is presented here. Let the motion of the massive relativistic string be defined as described in Subsection \ref{sec:alg}. In order to fragment the string and obtain the equations of motion for its daughters, one should perform the following actions:

\begin{enumerate}
    \item Use the formula (\ref{eq:prtndecay_fragm}) to sample the break time $\tau^*_{gj}$, $j=1,~\ldots,~2^g$ for each string with number $j$ of the current generation $g$.

    \item Sample two boundary values $l_{g(2j-1)}$, $l_{g(2j)}$ that determine the vanishing chunk of each string according to (\ref{eq:sigma_gap}), (\ref{eq:l_bound_gap}).
    
    \item Perform a Lorentz transformation to the center-of-mass frame of the considered mother string. The coordinates and velocity of the string can be obtained using (\ref{eq:ic_mod}).

    \item Calculate the masses of the daughter strings taking into account the angular momentum conservation using (\ref{eq:fragm_mass_mod}).

    \item Calculate the 4-momenta of the daughters with the formula (\ref{eq:str_energy_cm}), (\ref{eq:str_mom_cm}) and perform the Lorentz boost.

    \item Perform a Lorentz boost to the original frame of reference.

    \item Repeat steps 1 -- 6 until the fragmentation limit, determined by $l_{g(2j)} -l_{g(2j-1)} = 1$, is reached. 
\end{enumerate}

Note that this algorithm does not take into account the transition of strings to hadrons, which could significantly change the flow of the fragmentation.

An attentive reader has also noticed that in the defined scheme of fragmentation, one does not need to microscopically calculate the motion of the string. Only the integral properties of the string define the states with which the string-to-hadron transition algorithm should operate. This is natural as no measured hadron properties reflect the relative position of the quarks at the string ends in the coordinate space. The essence of the developed string model is portrayed only by the selection of masses of the daughter strings during the fragmentation.


\section{Discussion}
\label{sec:discuss}
A noticeably non-trivial picture has been achieved in the result of the development of the fragmentation model based on the classical Nambu-Goto string theory. We have started with the basic concepts that were already known for decades: the free string action (\ref{eq:NGaction}), the solution to the string equations of motion (\ref{eq:solution_std}) and the Virasoro conditions (\ref{eq:virasoro_cond}) for the initial data of the boundary value problem appearing as a consequence of using the orthonormal gauge (\ref{eq:ONG}). The problem of constructing strings from the arbitrary moving partons required developing the universal algorithm of calculating the motion of the string in $3+1$ dimensions. It turned out that the massive relativistic string should be considered as a rigidly rotating ``folded'' rod with a linear relation between the string spin and the mass squared. Upon that, it turned out that string breaking can occur only in a specific set of points that correspond to the ``joints'' of the ``folded'' rod-like strings. Moreover, to produce hadrons with momentum from the string at rest, a new mechanism of string breaking is required: the disappearance of the continuous set of the string points followed by the release of energy. While we have followed the strict mathematical principles of string theory, plenty of room remains for the discussion of possible modifications. In this Section some of them will be noted.

\paragraph{The FOEE(2)-string}
An obvious generalization of the proposed approach would be to consider the higher-order expansion (\ref{eq:foee_ic_fs}). All this time we have been interested only in the case of single non-zero term in the eigenfunction series. It is possible that, if more terms are added, the resulting string model will have very different limitations than the FOEE(1)-strings. Let us present the FOEE system derived for the case of two non-zero terms. The initial conditions for the string will be considered in the form:
\begin{equation}
    \label{eq:ic_foee2}
    \begin{aligned}
        v_{\mu}(\sigma) = a_{\mu} + b_{\mu} \cos(\sigma) + c_{\mu} \cos(2\sigma),\\
        \rho_{\mu}(\sigma) = d_{\mu} + e_{\mu} \cos(\sigma) + f_{\mu} \cos(2\sigma).
    \end{aligned}
\end{equation}
Then the Virasoro conditions accompanied by the conservation laws give the FOEE(2) system
\begin{equation}
    \label{eq:foee2_sys}
    \left\{
    \begin{aligned}
        cf = c^2 - 4f^2 = ce + 2bf = bc - 2ef &= 0\\
        be + \frac{4fP}{\kappa \pi} = b^2 - e^2 + \frac{2cP}{\kappa \pi}= bc + 2ef + \frac{2bP}{\kappa \pi} &= 0\\
        b^2 + c^2 + e^2 + 4f^2 + \frac{2M^2}{(\kappa \pi)^{2}} &= 0\\
        d_{\mu}P_{\nu} - d_{\nu}P_{\mu} + \frac{\kappa \pi}{2} \left( e_{\mu}b_{\nu} - e_{\nu}b_{\mu} + f_{\mu}c_{\nu} - f_{\nu}c_{\mu} \right) &= \mathcal{M}_{\mu\nu}
    \end{aligned}
    \right.
\end{equation}
The system (\ref{eq:foee2_sys}) consists of 15 equations and contains 20 unknown variables. The ability to define five more additional conditions provides some freedom of action, but, unfortunately, I was unable to solve this system even for the case of a string at rest. Neither any attempt to find a numerical solution was decisively successful. The non-linear nature of the FOEE system forces the complexity to grow very fast. It is not clear whether high-order FOEE systems have any solutions and if so, whether it is possible to solve them analytically.

\paragraph{The role of the Virasoro conditions}
Throughout the entire course of the development of the FOEE string model, Virasoro conditions were the main steering factor. Note that for the massless relativistic string the orthonormal gauge does not impose such restrictions, allowing the use of many different initial data functions. When requiring the mass of the string to be non-zero, a wide range of functions is forbidden, which led to the development of the Finite-Order Eigenfuction Expansion method. Due to the reasons explained previously, we were left with the case of the FOEE(1) $1_\nu^\xi$-string, i.e. the $\nu-1$ times ``folded'' rigid rod. The Virasoro conditions did not allow us to fragment the string at the arbitrary point. Instead, only the ``joints'' of the rod can be the break points. This, in turn, led to the need to create a new mechanism of string breaking with the release of energy. The Virasoro conditions also restrict us from changing the rotation frequency of the string, even when it is produced after fragmentation.

It is evident that these constraints will remain as long as the standard Nambu-Goto action is used. The resulting picture of the string fragmentation and dynamics seems too unnatural and artificial, though for the first time in the history of hadronization models it considers the role of the string spin in the fragmentation process. The reason for that might be the fact that the Nambu-Goto action describes the too mathematically pure object, rather than the realistic confined parton system. An investigation of alternative relativistic string actions is needed in order to determine whether similar constraints remain.

\paragraph{String with gluons}
Another important limitation of this model is the problem of the description of a string with gluons. Historically, it has been generally accepted to associate the (hard) gluons with ``kinks'' on a string stretched between $q$ and $\overline{q}$ \cite{jetmerge_1, jetmerge_2}. This follows the desire to describe the experimentally observed properties of jets \cite{jetprop}, most importantly the three-jet events that are interpreted as the result of the $qg\overline{q}$ string fragmentation. The ``kink'' in this approach represents the fact that the string does not stretch along the single line, but evolves in the first moments in time as a triangle shape with its vertices flying apart. This property of the string model is essential, and the question of generalization of the suggested FOEE-string model to the case of a string with gluons arises.

In the language of the Nambu-Goto string, the model of a ``string with kinks`` requires the segmentation of the string, i.e. for (at least) the initial conditions of the boundary value problem to be piece-wise defined. To be more precise, the segment-local momentum and/or angular momentum conservation should be imposed for at least the initial moment in time. In this way, the statement of the string motion problem would reflect the tendency to produce particles mostly in the directions defined by the momenta of $q$, $\overline{q}$ and $g$.

The issue is that a string defined by the initial conditions of the form
\begin{equation}
    \label{eq:gluonstr}
    \left\{
    \begin{aligned}
        v_{\mu} &= a_{1\mu} + b_{1\mu}\cos(\nu \sigma), \quad \rho_{\mu} = c_{1\mu} + d_{1\mu} \cos(\nu\sigma), \quad 0 \le \sigma \le \sigma_g,\\
        v_{\mu} &= a_{2\mu} + b_{2\mu}\cos(\nu \sigma), \quad \rho_{\mu} = c_{2\mu} + d_{2\mu} \cos(\nu\sigma), \quad \sigma_g < \sigma \le \pi,
    \end{aligned}
    \right.
\end{equation}
does not satisfy the Virasoro conditions. One might think that it is possible to solve the issue by a specific choice of the gluon point $\sigma_g$ as in the case of fragmentation. However, even if we choose $\sigma_g = s \pi/\nu$, where $s$ is an integer, $s<\nu$, there will still be infinitely many non-zero Fourier amplitudes due to the fact that the initial conditions are not orthogonal to the eigenfunction $u_0(\sigma)=1$. Clearly, any discontinuity in the initial conditions will result in the infinite number of terms in the Fourier series used to approximate such functions.

Overcoming this issue turns out to be quite a challenge. The Nambu-Goto action does not seem to provide the possibility of describing segmented strings. To avoid the restrictions imposed by the Virasoro conditions, a different string action may be required.

\paragraph{The Schwinger mechanism}
When the string breaks via $q\overline{q}$ pair production, the flavor is assigned to the newly created partons, and the correct description of the particle yields is impossible without the determination of the relative probabilities to produce different flavors. According to the Schwinger mechanism \cite{schwinger_1, schwinger_2, schwinger_3, schwinger_4}, in a constant (in space and time) electric field (with $\varepsilon$ the electric force) coupled to particles with mass $m$, the vacuum (the no-particle state) is unstable and decays according to an exponential law $\sim \exp(-p_{\perp})$. This mechanism was applied to the system confined by the QCD field \cite{schwinger_5, schwinger_6, schwinger_7, schwinger_8, schwinger_9, schwinger_10}. The pair of particles with mass (and, in general case, transverse momentum) cannot be produced in a single vertex. They must tunnel trough the region of the size $\sim m_{\perp}/\kappa$.  The probability
per unit of time and (in $1 + 1$ space-dimension) space can be calculated in the WKB approximation \cite{schwinger_7}, giving
\begin{equation}
    \label{eq:schwinger_pT}
    \frac{1}{\kappa} \frac{dP}{dp_{\perp}} \propto \exp{\left( -\frac{\pi m_{\perp}^2}{\kappa} \right)} = \exp{\left( -\frac{\pi m^2}{\kappa} \right)} \exp{\left( -\frac{\pi p_{\perp}^2}{\kappa}  \right)},
\end{equation}
and can be understood as the probability for a pair of particles with mass $m$ to tunnel out through a linear potential barrier. Since it is not clear what values of the quark masses to use, the relative probability for the flavor selection is tuned as a free parameter of the model.

The existing string models of hadronization assign the transverse momentum $\boldsymbol{p}_{\perp}$ sampled according to probability (\ref{eq:schwinger_pT}) to the end-points of the daughter strings. In the rest frame of the primary string, this is the only mechanism in these models giving the additional transverse momentum to the produced hadrons (we do not consider the collective effects now).

In the considered model of a $1_\nu^\xi$-string, the fragmentation occurs noticeably differently. Firstly, the parton pair ($q-\overline{q}$, $q-qq$ or $\overline{q}-\overline{qq}$) is not produced at a single point. The partons are separated by the entire length of the string (see figure \ref{fig:frag_sch_b}). This has no contradiction with the Nambu-Goto theory at all: as was said, the string has no inner forces of tension, so there are no limitations on the longitudinal inner string motion (and there is no determined longitudinal velocity of the string points either). Secondly, the conservation of the angular momentum forces the redistribution of the released energy, pushing the strings apart with the most natural direction selection being perpendicular to the string (rod) line. Thus, a mechanism to give the strings the transverse momentum $\boldsymbol{p}_{\perp}$ already exists within this model. If an additional momentum should be applied to the end point of the string, it can be done by changing the mass of the string as was done in (\ref{eq:fragm_mass_mod}), to match the new value of the angular momentum. Note that this will force one to change the masses of strings and recalculate their momenta using (\ref{eq:str_energy_cm}), (\ref{eq:str_mom_cm}). To take into account the additional momentum that is not aligned within the rotation plane of the string, the new rotation angles $\theta$, $\varphi$ in matrix (\ref{eq:rotation_mat}) should be defined.

As the length of the $1_{\nu}^{\xi}$-string is given by $l=2M/(\kappa \pi \nu)$, where $M$ is the mass of the string, one could say that the tunneling of the $q\overline{q}$ pair occurs through the same distance. Considering the Eq. (\ref{eq:schwinger_pT}), it might be interesting to study the influence of the invariant mass of the primary string and its spin on the relative probabilities of production of different flavors.

\paragraph{Prospects for hadron production}
The proposed model suggests an unfamiliar picture of string fragmentation, but opens some new possibilities in the description of hadron production. The most noticeable would be that one can now calculate the classical spin of the parton system modeled by a relativistic string right before the transition to the hadron state. This makes it possible to estimate the probabilities of vector/scalar particle production or, at least, to calculate the momentum ``kick'' for the neighboring string to conserve the angular momentum after the transition. In this way, it is possible, in principle, to achieve complete conservation of the angular momentum throughout the entire hadronization stage.

The second important feature is the existence of the natural limit to the fragmentation process determined by reaching the ``indivisible'' string state (i.e. the string defined on the interval $\sigma \in [r \pi / \nu, (r+1) \pi / \nu]$). If the mass of this string is too large for the string to be identified with the lightest hadron (pion), the heavier hadron of the same flavor content might be chosen instead, e.g., $\rho$-meson.

Another intriguing property of this model is the obvious difference between the fragmentation of the $ee$-, $pp$-, $pA$-, and $AA$-initiated system. The eigenvalue $\nu$ that determines the order of the non-zero harmonic of the string should be highly dependent on the impact parameter of the partons at the end points of the string. In this sense, the model suggests the natural transition from the physics of the short and low-spin strings in the $e^+e^-$ collisions to the long strings (in the transverse plane) with high rotation in the hadronic and nuclei collisions.

The obligatory introduction of the ``interstring'' interaction at the time of fragmentation determined by the released energy is also an essential mechanism. In principle, the higher the number of fragmentation stages occurred (high final-state multiplicity), the more energy is transferred into the additional momentum of the strings, which in turn might increase the average transverse momentum (if fragmentation occurs at the stage when the strings are oriented along the beam direction).

To give a quantitative analysis of the features considered in the model, a detailed simulation must be performed. This requires defining the exact mechanism of string-to-hadron transition, flavor selection, and rejection algorithms. The parton-level and preconfinement model used also greatly influences the results. Thus, such calculations go well beyond the scope of this article.

\section{Summary}
\label{sec:summary}
A brief review of the basics of the string model of hadronization is given. The restrictions imposed by the use of the orthonormal gauge in the classical Nambu-Goto string theory, called the Virasoro conditions, are reviewed. It is shown that the Virasoro conditions impose remarkably strong constraints on the functions that define the initial data for the boundary-value problem for the massive relativistic string motion. A wide class of functions does not satisfy these conditions, but only the eigenfunctions of the Sturm operator (represented by the $\cos(n\sigma)$, where $n$ is an integer for the string with free ends) being able to produce the finite and non-controversial Virasoro system.

A new method to define the initial conditions for the massive relativistic string is proposed. It is based on the Finite-Order Eigenfunction Expansion (FOEE) of the initial data functions $\rho_{\mu}(\sigma)$, $v_{\mu}(\sigma)$. The general form of the resulting system of Virasoro conditions is obtained. However, due to its complex non-linear structure, only the case of the first-order expansion is resolved (an FOEE(1)-string). An FOEE(1)-string considered in its center-of-mass frame turns out to be a rotating rigid rod with a fixed relation between its spin and mass. It is shown that both the Virasoro conditions and the conditions on the tangent vectors to the world sheet of the string are fulfilled.

As the string model should be able to describe the motion of the system of partons with arbitrary defined momenta, a method to calculate the motion of the FOEE-defined string in any frame of reference is developed. It is based on performing the Lorentz boost on the string described in the language of the distributed string quantities defined per unit $\sigma$. It is proven that this method of boosting the string does not break the Lorentz invariance of the theory. An important generalization of the FOEE(1)-string is considered: a string with a single eigenharmonic of arbitrary order (an $1_\nu^\xi$-string). It is shown that such a string can be interpreted as a ``folded'' rod with eigenvalue $\nu$ determining the number of times the string was folded. In this approach, the string spin is seen as a quantity with discrete spectrum. Using the $1_\nu^\xi$-string model, one can distinguish between the strings produced in $e^+e^-$ collisions, where no (significant) rotation momentum is expected, and the hadronic interactions, where the angular momentum of the system is defined by the impact parameter. The programmable algorithm to calculate the string motion is given.

The basics of the $1_\nu^\xi$-string fragmentation are developed. We see that the moment in time the string breaks can be sampled using a simple exponential formula. An arbitrary fragment of the string remains the time-like defined system. To provide a comprehensive description of the dynamics of the daughter string, the equations of motion of secondary strings are derived, as long as their solutions. The daughter strings are shown to be treated as strings, defined on the smaller $\sigma$-interval than their mother.

In order for the daughter strings to satisfy the Virasoro conditions, the string break point must be sampled among the discrete set of points that correspond to the ``joints'' of the ``folded'' string (rod). This is the only option allowed. The simple algorithm to sample the break point of the string of any given generation is described. The resulting equations of motion of secondary strings are derived. We see that the fragmentation process must stop at the string defined between the two nearest ``joints''. In this way, a natural limit to the string fragmentation is introduced, and, thus, the issue with infinite fragmentation is avoided.

A proof that all daughter strings satisfy the Virasoro conditions is given. An important relation between the mass and spin of the daughter strings is obtained. Although the primary $1_\nu^\xi$-string with large mass can have a moderate spin value (so the slope of the line $J(M^2)$ is small), the relation between the spin of the daughter string $J$ and its mass $M$ corresponds to the Regge trajectory with $J=M^2/(2\kappa \pi)$ in the limit of extreme fragmentation.

As fragmentation by a single point would result in the production of hadrons at rest, a new mechanism of string fragmentation is proposed. The fragmentation of the string is described as the result of vanishing of the continuous set of points (a chunk) of the string allowing the release of energy. To conserve the total angular momentum of the system, the mechanism of redistributing the properties of the vanished chunk is introduced. The angular momentum of the daughter strings is corrected to conserve the total value, and their masses are changed correspondingly. With their masses known, the 2-body decay laws can be used to calculate the additional momentum the two fragments gain after the mother string breaks. The resulting algorithm to calculate the fragmentation of the massive relativistic string is also formulated.

\appendix
\section{Derivation of the solution to the problem for the motion of the relativistic string}
\label{sec:appA}

The formulation of the Cauchy problem is as follows:
\begin{equation}
    \label{eq:ACauchyProb}
    \begin{aligned}
         \ddot{x}_{\mu}-x^{\prime\prime}_{\mu} &= 0,\quad \sigma \in\left[0, \pi\right], \quad \tau>0, \quad \mu=0,~\ldots,~3;\\
         x^{\prime}_{\mu}(\tau,0) & =x^{\prime}_{\mu}(\tau,\pi) = 0;\\
         x_{\mu}(0,\sigma) & = \rho_{\mu}(\sigma), \quad \dot{x}_{\mu}(\sigma) = v_{\mu}(\sigma),
    \end{aligned}
\end{equation}
where
\begin{equation*}
    \dot{x}_{\mu}(\tau,\sigma) \equiv \frac{\partial x_{\mu}(\tau,\sigma)}{\partial\tau},\quad x^{\prime}_{\mu}(\tau,\sigma) \equiv \frac{\partial x_{\mu}(\tau,\sigma)}{\partial\sigma}.
\end{equation*}
Let us express the solution as $x_{\mu}(\tau,\sigma)=T_{\mu}(\tau)u(\sigma)$. Substitution of the solution in this form into differential equation yields
\begin{equation}
    \label{eq:Asep_var}
    \frac{\ddot{T}_{\mu}(\tau)}{T_{\mu}(\tau)} = \frac{u^{\prime\prime}(\sigma)}{u(\sigma)} = -{\omega}^2,
\end{equation}
where $\omega$ is a constant. Eq. (\ref{eq:Asep_var}) gives a differential equation for the function $u(\sigma)$. When adding the boundary conditions of the problem (\ref{eq:ACauchyProb}) one obtains what is called a Sturm-Liouville problem on eigenfunction $u(\sigma)$ that corresponds to eigenvalue $\omega$:
\begin{equation}
    \label{eq:ASturmProb}
    \begin{aligned}
         u^{\prime\prime}(\sigma) + \omega^2 u(\sigma) &= 0,\quad \sigma \in\left[0, \pi\right];\\
         u^{\prime}(0) = u^{\prime}(\pi) & = 0.\\
    \end{aligned}
\end{equation}
General solution to the problem (\ref{eq:ASturmProb}) can be expressed as:
\begin{equation}
    \label{eq:Aufunc_gen}
    u_{n}(\sigma) = A_n\cos(\omega_n \sigma)+B_n \sin(\omega_n \sigma).
\end{equation}
With boundary conditions applied to Eq. (\ref{eq:Aufunc_gen}) one obtains:
\begin{equation}
    \label{eq:Aufunc}
    \begin{aligned}
        \omega_n &= n,\quad n=1,~2,~\ldots~,\\
        u_{n}(\sigma) & = \cos(\omega_n \sigma).\\
    \end{aligned}
\end{equation}
The case of $\omega_0=0$ should also be considered: $u_0(\sigma) \equiv 1$.
The essential property of eigenfunctions (\ref{eq:Aufunc}) is their orthogonality:
\begin{equation}
    \label{eq:Aef_ort}
    \begin{aligned}
        \int_{0}^{\pi}u_n(\sigma)u_m(\sigma)d\sigma &= \frac{\pi}{2} \delta_{nm} \equiv {||u_n||}^2\delta_{nm},\quad n \ne 0,\\
        \int_{0}^{\pi}u_n(\sigma)u_0(\sigma)d\sigma &= \pi \delta_{n0}\equiv{||u_0||}^2\delta_{n0}.
    \end{aligned}
\end{equation}
Here, $||u_n||$ is the norm of the function $u_n(\sigma)$ and $\delta_{nm}$ is the Kronecker delta. The general solution for the function $T_{\mu}(\tau)$ can be written the following way:
\begin{equation}
    \label{eq:ATfunc}
    \begin{aligned}
        T_{n\mu}(\tau) &= C_{n\mu} \sin(\omega_n\tau) + D_{n\mu}\cos(\omega_n \tau),\quad \omega_n \equiv n, \quad n=1,~2,~\ldots,\\
        T_{0\mu}(\tau) &= C_{0\mu} \tau + D_{0\mu}.
    \end{aligned}
\end{equation}

The solution to the initial problem can be expressed as a Fourier series over partial solutions (\ref{eq:Aufunc}), (\ref{eq:ATfunc}):
\begin{equation*}
    x_{\mu}(\tau, \sigma) = T_{0\mu}(\tau)u_0(\sigma) + \sum_{n=1}^{+\infty}T_{n\mu}(\tau)u_n(\sigma).
\end{equation*}
Now substituting this expression into the first initial condition, one obtains the formula for coefficients $D_{n\mu}$:
\begin{equation}
    \label{eq:ADcoef}
    \begin{aligned}
        D_{0\mu} &= \frac{\int_{0}^{\pi}\rho_{\mu}(\sigma)u_0(\sigma)d\sigma}{{||u_0||}^2} = \frac{1}{\pi} \int_{0}^{\pi}\rho_{\mu}(\sigma)d\sigma,\\
        D_{n\mu} &= \frac{\int_{0}^{\pi}\rho_{\mu}(\sigma)u_n(\sigma)d\sigma}{{||u_n||}^2} = \frac{2}{\pi} \int_{0}^{\pi}\rho_{\mu}(\sigma) \cos(n\sigma)d\sigma, \quad n \ne 0.\\
    \end{aligned}
\end{equation}
Second initial condition yields expressions for coefficients $C_{n\mu}$:
\begin{equation}
    \label{eq:ACcoef}
    \begin{aligned}
        C_{0\mu} &= \frac{\int_{0}^{\pi} v_{\mu}(\sigma)u_0(\sigma)d\sigma}{{||u_0||}^2} = \frac{1}{\pi} \int_{0}^{\pi} v_{\mu}(\sigma)d\sigma,\\
        C_{n\mu} &= \frac{\int_{0}^{\pi} v_{\mu}(\sigma)u_n(\sigma)d\sigma}{{\omega_n||u_n||}^2} = \frac{2}{n\pi} \int_{0}^{\pi} v_{\mu}(\sigma) \cos(n\sigma)d\sigma, \quad n \ne 0.\\
    \end{aligned}
\end{equation}
Note that expressions in the middle are valid for arbitrary eigenfunctions and eigenvalues that might appear if different boundary conditions are used for the string.

The general solution takes the form:
\begin{equation}
    \label{eq:Ax_long}
    \begin{aligned}
        x_{\mu}(\tau, \sigma) &= \frac{\tau}{\pi} \int_{0}^{\pi} v_{\mu}(\sigma)d\sigma + \frac{1}{\pi} \int_{0}^{\pi}\rho_{\mu}(\sigma)d\sigma\\ 
        &+ \sum_{n=1}^{+\infty} \left[ \frac{\sin(n\tau)}{\pi n} \int_{0}^{\pi} v_{\mu}(\lambda) \cos(n\lambda)d\lambda + \frac{\cos(n\tau)}{\pi}\int_{0}^{\pi} \rho_{\mu}(\lambda) \cos(n\lambda)d\lambda\right] \cos(n\sigma).
    \end{aligned}
\end{equation}
To shorten Eq. (\ref{eq:Ax_long}), it is useful to use exponential expressions for $\sin(n\tau)$ and $\cos(n\tau)$ and to take into account that the integrands in (\ref{eq:Ax_long}) are independent of the sign of $n$. Thus, one can obtain:
\begin{equation}
    \label{eq:Ax_shorter}
    \begin{aligned}
        x_{\mu}(\tau, \sigma) &= \frac{\tau}{\pi} \int_{0}^{\pi} v_{\mu}(\sigma)d\sigma + \frac{1}{\pi} \int_{0}^{\pi}\rho_{\mu}(\sigma)d\sigma\\ 
        &+ \frac{i}{\pi}\sum_{n \ne 0} \frac{e^{-in\tau}}{n} \cos(n\sigma) \int_{0}^{\pi} \left[  v_{\mu}(\lambda) - in \rho_{\mu}(\sigma) \right] \cos(n\lambda) d\lambda.
    \end{aligned}
\end{equation}
The only step left to be taken is to introduce the notations:
\begin{equation*}
    P_{\mu} \equiv  \kappa \int_{0}^{\pi} v_{\mu}(\sigma) d\sigma, \quad Q_{\mu} \equiv  \frac{1}{\pi} \int_{0}^{\pi} \rho_{\mu}(\sigma) d\sigma,
\end{equation*}
\begin{equation*}
    \alpha_{n\mu} = \sqrt{\frac{\kappa}{\pi}} \int_{0}^{\pi}\left[ v_{\mu}(\sigma) - in \rho_{\mu}(\sigma) \right]  \cos(n\sigma) d\sigma, \quad n \ne 0,\\ 
\end{equation*}
in order to shape the formula for string coordinates into the standard form:
\begin{equation}
    \label{eq:Asolution_std}
    x_{\mu}(\tau,\sigma) = Q_{\mu} + \frac{P_{\mu}\tau}{\kappa \pi} + \frac{i}{\sqrt{\kappa \pi}} \sum_{n \ne 0} e^{-in\tau}\frac{\alpha_{n\mu}}{n}\cos(n\sigma).
\end{equation}

\acknowledgments
The author thanks Professor Anatoly Petrukhin for the comprehensive assistance, valuable advice, and the opportunity to carry out this research. The work was performed with the support of the Ministry of Science and Higher Education of the Russian Federation, project ``Fundamental and applied research of cosmic rays'', No. FSWU-2023-0068.

 \bibliographystyle{JHEP}
 \bibliography{biblio}

\end{document}